\documentclass[aps,prd,reprint,superscriptaddress,showpacs,floatfix,nofootinbib]{revtex4-2}  % for review and submission

\usepackage[hidelinks,draft=false]{hyperref}
\usepackage{import}

\usepackage{graphicx}
\usepackage{amsmath,amssymb,graphics,setspace}
\usepackage{color}
\usepackage[english]{babel}
\usepackage{url,bm}
\usepackage{amsfonts}
\usepackage[separate-uncertainty = true]{siunitx}
\usepackage{dcolumn}
\usepackage{appendix}
\usepackage{lmodern}
\usepackage[draft=false]{microtype}
\usepackage{esdiff}
\usepackage[table,dvipsnames]{xcolor}
\usepackage[babel]{csquotes}
\usepackage{textcomp}
\usepackage{listings}
\usepackage{enumitem}
\usepackage{lipsum}
\usepackage{tabularx}
\usepackage{makecell}
\usepackage{soul}
\usepackage{enumitem}
\usepackage{afterpage}
\usepackage{float}
\usepackage{wasysym}

\usepackage{booktabs, tabularx, wrapfig}

\usepackage{siunitx}

\usepackage{natbib}
\newcommand{\mathsym}[1]{{}}
\newcommand{\unicode}[1]{{}}

\newcommand{\etr}{\mathrm{etr}}
\newcommand{\cw}{\mathcal{CW}}
\newcommand{\bSig}{\bm{\Sigma}}
\newcommand{\bS}{\bm{S}}
\newcommand{\bW}{\bm{W}}

\newcommand{\bU}{\bm{U}}

\newcommand{\bQ}{\bm{Q}}
\newcommand{\bPi}{\bm{\Pi}}
\newcommand{\balp}{\bm{\alpha}}

\renewcommand{\Re}{\operatorname{Re}}
\renewcommand{\Im}{\operatorname{Im}}

\definecolor{HotPink}{RGB}{255 ,0,128}
\definecolor{cardinal}{rgb}{0.77, 0.12, 0.23}
\definecolor{Gorange}{RGB}{255,80,0}
\definecolor{lightgray}{gray}{0.85}

\begin{document}

\title [mode = title]{Precision spectral estimation at sub-Hz frequencies:\texorpdfstring{\\}{ }Closed-form posteriors and  Bayesian noise projection}

\begin{abstract}
We consider the problem of estimating cross-spectral quantities in the low-frequency regime, where long observation times limit averaging over large ensembles of periodograms, thereby preventing the use of approximate Gaussian statistics. This case is relevant for precision low-frequency gravitational experiments such as LISA and LISA Pathfinder.\\
We present a Bayesian method for estimating spectral quantities in multivariate Gaussian time series. The approach, based on periodograms and Wishart statistics, yields closed-form expressions at any given frequency for the marginal posterior distributions of the individual power spectral densities, the pairwise coherence, and the multiple coherence, as well as for the joint posterior distribution of the full cross-spectral density matrix. In the context of noise projection---where one series is modeled as a linear combination of filtered versions of the others, plus a background component---the method also provides closed-form posteriors for both the susceptibilities, i.e., the filter transfer functions, and the power spectral density of the background.
We apply the method to data from the LISA Pathfinder mission, showing effective decorrelation of temperature-induced acceleration noise and reliable estimation of its coupling coefficient. 
\end{abstract}

\newcommand{\ORCIDiD}[1]{\href{https://orcid.org/#1}{\includegraphics[width=2ex]{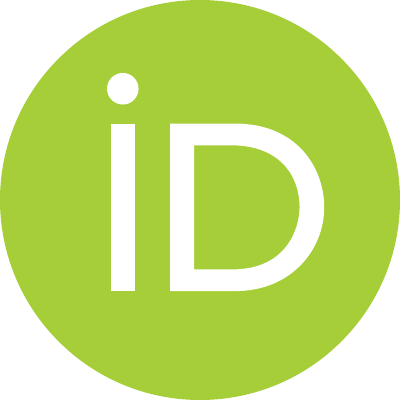}}}
\author{Lorenzo~Sala~\ORCIDiD{0000-0002-2682-8274}}\affiliation{\addressunitn}\affiliation{\addresstifpa}
\author{Stefano~Vitale~\ORCIDiD{0000-0002-2427-8918}}\affiliation{\addressunitn}
\def\addressunitn{Dipartimento di Fisica, Universit\`a di Trento, 38123 Trento, Italy}
\def\addresstifpa{Trento Institute for Fundamental Physics and Application / INFN, 38123 Trento, Italy}
\date{September 9, 2026}

% \pacs{02.50.Tt, 02.50.Sk, 07.05.Kf, 04.80.Nn}
\maketitle

% \interfootnotelinepenalty=10000

\section{Introduction} 
\label{sec:intro}

Spectral analysis based on Welch’s method \cite{Welch1967,Papoulis} is the standard approach for noise characterization. In this framework, power spectral densities (PSD) and cross-spectral densities (CPSD) of multivariate noise time series are estimated from properly normalized discrete Fourier transforms, known as periodograms.\\
To improve the precision of these estimates and to quantify their uncertainty, Welch’s method partitions the time series into $M$ equal length, possibly overlapping segments, yielding $M$ independent (or very weakly correlated) periodogram samples. In the usual frequentist approach, PSD and CPSD are obtained by averaging these samples, with uncertainties scaling with the standard deviation of the mean. This procedure relies on the central limit theorem, which ensures that the averaged estimates rapidly approach Gaussian statistics, enabling the construction of confidence intervals.\\
However, spectral resolution decreases with decreasing $M$, and in many applications, particularly those at very low frequency, $M$ is inevitably small, sometimes even $M = 1$. As $M$ decreases, the Gaussian approximation for the statistics of the average
becomes increasingly inaccurate, leading, for instance, to paradoxical outcomes such as non-negligible probabilities for negative PSD values. At $M = 1$, the frequentist approach becomes entirely inapplicable. By contrast, a Bayesian framework does not suffer from these limitations and remains well defined even in this regime.

In the first part of this paper, we derive the posterior distributions for many spectral quantities, under a Bayesian framework, focusing in particular on the case where just a few periodograms are available, that is, far from Gaussianity.

In the second part of this paper, starting from Sec.~\ref{sec:tsdecorr}, we address a closely related problem: noise projection and time-series decorrelation. Specifically, we consider the case in which one time series is modeled as a linear combination of filtered versions of other series, plus an additional background component. We develop a Bayesian method to infer both the residual noise properties (after decorrelation) and the coupling coefficients. Particular attention is devoted to the underlying probability distributions, which can significantly deviate from Gaussianity when only a small number of periodograms is available. We provide a detailed technical discussion of the method, validate it through numerical simulations, and demonstrate its application to real data.

We encountered a need for these methods while analyzing data from the LISA Pathfinder (LPF) mission \cite{PhysRevD.110.042004_LPFnoiseperf2024,PhysRevLett.120.061101_beyond,PhysRevLett.116.231101_subfemtog}. The mission’s objective was to precisely measure the noise spectrum of force disturbances acting on two nominally freely falling test masses in space, reaching acceleration levels as low as a few \si{\femto\meter\,\second^{-2}/\hertz^{1/2}}, and frequencies down to approximately \SI{20}{\micro\hertz}. 
Unlike typical spectral estimation applications, where the goal is to extract a signal from a noisy stochastic process, LPF aimed to measure the noise itself with the highest possible accuracy, particularly at the lowest achievable frequencies.  \\
While reviewing the literature for a consistent and practical Bayesian approach suited to our needs, we found that the fundamental principles had long been established. However, we could not find a detailed, practical method applicable to our data processing, leading us to develop one independently.  \\
We applied this method to the data analysis of the LPF mission in Refs.~\cite{PhysRevD.110.042004_LPFnoiseperf2024,PhysRevD.109.102009_LPFactuation2024}, and briefly summarized its main features in the appendices of the second paper. Here, we provide a detailed description of the method, discussing its foundations, deriving key procedures rigorously, and presenting quantitative evidence of its validity through numerical simulations.

The paper is organized as follows: in Sec.~\ref{sec:CPSDstats} we define the key experimental quantities of multivariate time series and derive their likelihood under the Gaussian data hypothesis; in Sec.~\ref{sec:CPSD_TH_stats_a} we build the Bayesian posteriors for all the related spectral quantities.
In Sec.~\ref{sec:tsdecorr} we discuss the case of noise projection, that is, the case where one series is modeled as a linear combination of filtered versions of the others, plus a background component of which one wants to estimate the spectrum; in Sec.~\ref{sec:simulationsapplications} we apply our method to simulations and real data. Finally, in Sec.~\ref{sec:conclusions} we give some concluding remarks.

The method presented in this paper is available as a Python package at \cite{Sala_noisefinder_2026}.

\section{Periodograms, key periodogram functions and their likelihood} 
\label{sec:CPSDstats}
In this section, we recall a few basic concepts and results that we will use in our Bayesian inference method. We assume that data are Gaussian for the purpose of the our analysis, something we found to be quantitatively true in the case of LPF \cite{PhysRevD.110.042004_LPFnoiseperf2024}.

\subsection{Basic definitions and nomenclature} 
\label{sec:definitions}
In our approach, we assume that we have acquired, synchronously and with sampling time $T$, the time series of $p$ real, stationary, Gaussian, zero-mean stochastic processes. We call $x_i[n]$, with $1\le i \le p$, the sample of the $i$th series taken at time $t=n T$. \\
The joint statistics of these samples is fully contained in the mean values of their products
\begin{equation}
    \langle x_i[n]x_j[m]\rangle=\int_{-\infty}^\infty\Sigma_{i,j}(f)e^{i2\pi(m-n)fT} {d}f,
\end{equation}
 and then in the  Hermitian positive definite matrix $\bSig(f)$, with elements $\Sigma_{i,j}(f)$, which, by definition, is the joint two-sided power cross-spectral density CPSD matrix of the $p$ stochastic processes at frequency $f$. Note that the diagonal element $\Sigma_{i,i}(f)$ is the power spectral density PSD of the process $x_i(t)$, while the off-diagonal $\Sigma_{i,j}(f)$ is the pair CPSD of $x_i(t)$ and $x_j(t)$.\\
Our goal is to infer each element of $\bSig(f)$ independently at each frequency, without assuming any functional dependence on $f$.

The main tool for such an inference is the periodogram $X_i[k]$ calculated over an $N$-long segment of the multivariate time series. $X_i[k]$ is defined as:
\begin{equation}
\label{eq:periodogram}
    X_i[k] = \sqrt{\frac{T}{N}} \sum_{n=0}^{N-1} x_i[n] \, w[n] \, e^{-2\pi i k  n/N}
\end{equation}
with $0\le k \le N-1$ an integer, and $w[n]$  the coefficients of a suitable tapering window. Since $x_i[N-k]=x_i^*[k]$, only the first $\lfloor N/2\rfloor+1$ of these coefficients carry independent information. $X_i[k]$ is Gaussian, complex, and zero-mean.  Key for the inference are the complex mean values:
\begin{equation}
\label{eq:wind1}
\begin{split}
    &\langle X_i[k]X_j^*[k']\rangle=\frac{T}{N}\int_{-\infty}^\infty\Sigma_{i,j}(f)\times\\
    &\times \tilde{w} \left(\frac{2\pi}{N}k-2\pi f T\right)\tilde{w}^* \left(\frac{2\pi}{N}k'-2\pi f T\right) \,df
    \end{split}
\end{equation}
with $\tilde{w}(\phi)=\sum_{n=0}^{N-1}w[n]e^{-i\phi n}$ the Fourier sequence transform of the tapering window $w[n]$. 
If, ideally, one could choose $w[n]$ such that
\begin{equation}
\begin{split}
\tilde{w} &\left(\frac{2\pi}{N}k-2\pi f T\right)\tilde{w}^* \left(\frac{2\pi}{N}k'-2\pi f T\right)=\\
&= \frac{N}{T} \delta_{kk'}\delta\left(\frac{2\pi}{N}k-2\pi f T\right) ,
\end{split}
\end{equation}
with $\delta_{kk'}$ the Kronecker delta of $k$ and $k'$, and $\delta(\phi)$ the Dirac delta of $\phi$, then $X_i[k]X_j^*[k]$ would be an unbiased estimator of $\Sigma_{ij}\left(f=k/(NT)\right)$, and $X_i[k]$ would be independent of $X_j^*[k']$ if $k\neq k'$.

In reality, $\tilde{w}(\phi)$ is a $2\pi$-periodic function with a central lobe at $\phi=0$, and a sequence of strongly suppressed side lobes. We assume that the aliasing deriving from the periodicity of $\tilde{w}(\phi)$ has been made negligible by properly choosing a short enough sampling time $T$. \footnote{If this assumption does not hold, then $\Sigma_{i,j}$ in Eq.~\eqref{eq:wind1} should instead be interpreted as the discrete-time CPSD of the discrete stochastic process obtained by sampling the original continuous-time process. All statistical conclusions presented in the remainder of the paper remain unchanged. However, if these discrete-time quantities are used to estimate their continuous-time counterparts, the resulting estimates will be systematically biased due to aliasing.}\\
The width of the central lobe depends on the choice of the window, but is always of the form $\pm m (2\pi/N)$, with $m$ a small integer. Thus, if $\lvert k-k'\rvert >m$, then $X_i[k]$ and  $X_j^*[k']$ may be treated as independent within a reasonable accuracy\footnote{In this paper, when showing plots with spaced frequency points, we always choose $k$ and $k'$ so that adjacent frequencies exhibit a controlled correlation in the range of 10\% to 30\%, while correlations between second-nearest neighbors are negligible.}. Then, from Eq.~\eqref{eq:wind1},
\begin{equation}
\label{eq:wind2}
    \langle X_i[k]X_j^*[k]\rangle\simeq\int_{-\frac{1}{2T}}^{\frac{1}{2T}}\Sigma_{i,j}(f)G\left(f-\frac{k}{NT}\right)df
\end{equation}
with $G(f) = \left\lvert \tilde{w}(2\pi f T) \right\rvert^2$. Note that $w[n]$ is always normalized such that  $\int_{-\frac{1}{2T}}^{\frac{1}{2T}}G\left(f-\frac{k}{NT}\right) df=1$.\\
Thus, $X_i[k]X_j^*[k]$ is an estimator of $\Sigma_{i,j}(f)$ averaged over the  band  $f=(k \pm m)/(N T)$.  \\ 
From now on, unless otherwise specified, we indicate with $\bSig(f)$ this averaged version of the CPSD matrix.

As usual in statistics, the precision of the estimator can be increased by averaging over repeated measurements. To this aim, Welch's method prescribes to split the available time series into $M$ segments\footnote{These segments do not need to be disjoint. It has been shown that, as the window $w[n]$ tapers their ends, some overlap between adjoining segments does not significantly change the statistical properties of the derived quantities with respect to the case of disjoint segments \cite{Welch1967}.}, each of length $N$, average over them, and use the observed CPSD matrix $\bPi[k]$, with elements
\begin{equation}
    \label{eq:CPSD}
    \Pi_{ij}[k] = \frac{1}{M} \sum_{\ell=1}^M X_{i,(\ell)}[k] X_{j,(\ell)}^*[k],
\end{equation}
as an estimator of $\bSig(f=k/(NT))$.\\
Thus, in summary,  the matrices $\bPi[k]$, with $k\in[0, 2m,$ $4m, 6m,...,\lfloor N/4m\rfloor]$ are independent estimators of the matrices $\bSig(f)$ with $f \in [0, 2m, 4m, 6m,...,\lfloor N/4m\rfloor]/(NT)$, with a spectral resolution of $\pm m /(NT)$.\\
One can show that $\bPi(k)$, which is Hermitian, is positive definite only if  $M\ge p$. As the positive definiteness is mandatory for an estimator of $\bSig(f=k/(NT))$, the minimum number of periodograms one should average on is $p$.

Some common applications deviate from the spectral estimator with evenly spaced frequencies described above \cite{PhysRevD.110.042004_LPFnoiseperf2024,heinzel-trobs-spectrum-estimation, PSDnew}.  In those applications, at each frequency of interest $f$, one adjusts $M$, and then $N$, according to some averaging optimization criterion, and picks just one matrix  $\bPi[k]$, with $k$ selected such that $f=k/(NT)$.  $\bPi[k]$ is then an estimate of the CPSD at the frequency $f$. This procedure is repeated at each frequency of interest.\\
As $N$ depends on the frequency, the width of the spectral window is no longer frequency-independent, as it is in the case of uniform spacing. As a consequence, the independence of the CPSD estimators at different frequencies must, in principle, be established frequency by frequency. An assessment of the methods in Refs.~\cite{PhysRevD.110.042004_LPFnoiseperf2024} and~\cite{PSDnew} shows that, for at least some choices of parameters, the estimators at different frequencies can be considered practically independent. We will use these methods in several examples presented in the remainder of the paper.
 
Some functions of $\bPi[k]$ that we will also use in the following are:
\begin{enumerate}[label=(\roman*)]
    \item the measured  magnitude-squared cross-coherence (MSC), between the $i$th and $j$th time series
\begin{equation}
    \left\lvert\hat{\rho}_{ij}[k]\right\rvert^2 = \left\lvert\frac{\Pi_{ij}[k]}{\Pi_{ii}^{1/2}[k] \, \Pi_{jj}^{1/2}[k]}\right\rvert^2;
\end{equation}
a useful diagnostics of the possible linear correlation between the two underlying processes;
\item the multiple coherence \cite{goodman_statistical_1963}, a useful generalization of $ \left\lvert\hat{\rho}_{ij}[k]\right\rvert^2 $ to the case of multiple series, 
\begin{equation}
    \label{eq:R2_defi}
    \hat R^2[k] = 1 - \left( \Pi_{11}^{\vphantom{-1}}[k] \, \Pi_{11}^{-1}[k]\right)^{-1},
\end{equation}
with $\Pi_{i,j}^{-1}$ the elements of the inverse $\bPi^{-1}$ of $\bPi$. This is used as a diagnostic of how much of the noise power in $x_1[n]$ is due to its correlation to the remaining series.
\item The Schur complement of any sub-block $\bm{C}$ in the decomposition of the  Hermitian matrix $\bPi[k]$ as
\begin{equation}
    \label{eq:schurblocks2}
    \bPi[k] = \left(\begin{matrix}
    \bm{A} & \bm{B} \\
    \bm{B}^\dagger & \bm{C} \\
    \end{matrix}\right)
\end{equation}
Here $\bm{A}$ is a $q\times q$ matrix, $\bm{C}$ is $r\times r$, $\bm{B}$ is $q\times r$, and $q+r=p$. The Schur complement of the block $\bm{C}$ in $\bPi[k]$, is:
\begin{equation}
    \bPi[k]/\bm{C} \equiv \bm{A} - \bm{B} \bm{A}^{-1} \bm{B}^\dagger
\end{equation}

\end{enumerate}

We discuss in the following section the sampling distributions of all these quantities.

\subsection{Sampling distribution of the  CPSD matrix}
\label{sec:CWdistrib}
Reference~\cite{goodman_statistical_1963} shows that the joint sampling distribution of the elements of the matrix  $\bW=M\bPi$, conditional to the theoretical CPSD matrix $\bSig$, is a complex Wishart distribution, with probability density function (PDF):
\begin{equation}
\label{eq:wishart}
    p\left(\bW\big\vert\bSig,M \right) = \frac{\left|\bW \right|^{M-p}}{\widetilde{\Gamma}_p(M) \left| \bSig \right|^{M}} \ \etr \left[- \bSig^{-1} \bW\right]
\end{equation}
Here, $\left|\cdot\right|$ is the determinant, $\etr$ the exponential trace $\etr(\cdot)=\exp\left(\mathrm{tr}\left(\cdot\right)\right)$, and $\widetilde{\Gamma}_p(M)$ is the multivariate complex gamma function:
\[
\widetilde{\Gamma}_p(M) = \pi^{\frac{1}{2} p(p-1)} \prod_{i=1}^{p} \Gamma(M-i+1)
\]
Note that we have dropped, for clarity, the explicit dependence of all quantities on frequency.\\
We denote this distribution\footnote{We use the symbolic expression $a \sim A$ to indicate that a random variable $a$ is distributed according to a distribution $A$.\\ Thus, $\bW\sim\cw(\bSig,M)$.} with  $\cw(\bSig,M)$. As expected, $\cw(\bSig,M)$ is defined only if $M\geq p$,  that is, if $\bW$ is positive definite.\\
We use the conditional probability in Eq.~\eqref{eq:wishart} as the likelihood function for the inference of $\bSig$.

\subsection{Sampling distribution of derived quantities}
\label{sec:CPSD_EXP_stats}
The complex Wishart distribution describes the joint probability of all the elements of the matrix $\bW$. Starting from that, and employing its mathematical properties \cite{goodman_statistical_1963, nagargupta_wishart}, we give the sampling distributions of some derived quantities that we will use as likelihood functions  for the  Bayesian inference of the relative quantities.

\paragraph{Power spectral density.} 
%For $p=1$, that is, in case of a single univariate stochastic process,  calling  $\Pi$ the only element of $\bPi$, and $S$ the PDF of the process and only element of $\bSig$, the PDF in Eq.~\eqref{eq:wishart}  reduces to
For $p=1$, that is, in the case of a single univariate stochastic process, let $W$, $\Pi$, and $S$ denote the only elements of $\mathbf{W}$, $\boldsymbol{\Pi}$, and $\boldsymbol{\Sigma}$, respectively. Since $\mathbf{W}=M\boldsymbol{\Pi}$, it follows that $W=M\Pi$. The PDF in Eq.~\eqref{eq:wishart} then reduces to
\begin{equation}
\label{eq:directPSD}
    p\left(M\Pi\big\vert S \right) = \frac{ (M \Pi)^{M-1}}{\Gamma(M) S^M} \ e^{- M \Pi/S}
\end{equation}
This means that $\Pi\sim \Gamma(M,S/M)$, with $\Gamma(M,S/M)$ the gamma distribution with shape parameter $M$  and scale parameter $S/M$. Equivalently, Eq.~\eqref{eq:directPSD} implies that $2M\Pi/S$ is chi-square distributed with $2M$ degrees of freedom. Note that this result is also obtained by calculating the marginal distribution of any of the diagonal elements of $\bPi$ from the joint PDF in Eq.~\eqref{eq:wishart}. 

\paragraph{Magnitude squared coherence.} Defining the theoretical $\rho_{ij}$
\begin{equation}
   \rho_{ij}=\frac{\Sigma_{i,j}}{\Sigma_{i,i}^{1/2}\Sigma_{j,j}^{1/2}}, 
\end{equation}
the sampling distribution of the MSC, $|\hat\rho_{ij}|^2$ is \cite{goodman_statistical_1963,carter_estimation_1973}: 
\begin{align}
\label{eq:directRho}
    p(|\hat{\rho}|^2\big\vert|\rho|^2) = & (M-1)(1-|\hat{\rho}|^2)^{M-2}(1-|\rho|^2)^{M} \nonumber\\
    &\times \ {}_2F_1(M,M,1,|\hat{\rho}|^2\,|\rho|^2)
\end{align}
where ${}_2F_1$ represents  Gauss' hypergeometric function. \\
Note that this distribution only holds for $M>1$ as,  notoriously, when $M=1$, $|\hat\rho|^2=1$ holds exactly. More in general, for low values of $M$,  $p(|\hat{\rho}|^2\big\vert|\rho|^2)$ carries a significant bias toward $|\hat{\rho}|^2>|\rho|^2$.

\paragraph{Multiple coherence.} 
Similarly to the case of MSC, defining the theoretical $R^2$:
\begin{equation}
    R^2=1-(\Sigma_{1 1}\Sigma_{11}^{-1})^{-1}
\end{equation}
with $\Sigma_{i,j}^{-1}$ the elements of $\bSig^{-1}$, 
the PDF of the sample multiple coherence $\hat{R}^2$  is  \cite{goodman_statistical_1963}:
\begin{align}
\label{eq:directR2}
    p(\hat{R}^2\big\vert R^2) = & \frac{\Gamma(M)}{\Gamma(p-1)\Gamma(M-p+1)}\, (\hat{R}^2)^{p-2}(1-\hat{R}^2)^{M-p} \nonumber\\
    &\times (1-R^2)^M \,  {}_2F_1(M,M,p-1,R^2\hat{R}^2)
\end{align}
with $M\ge p$. In the 2-D case, the multiple coherence and the MSC coincide.

\paragraph{Schur complement} Finally,  if  $\bW$ is  decomposed as:
\label{par:schurwishart}
\begin{equation}
    \label{eq:schurblocks}
    \bW = \left(\begin{matrix}
    \bm{A} & \bm{B} \\
    \bm{B}^\dagger & \bm{C} \\
    \end{matrix}\right)
\end{equation}
with  $\bm{C}$ an $r\times r$ matrix, then  the Schur complement of  $\bm{C}$, 
\begin{equation}
    \bW/\bm{C} \equiv \bm{A} - \bm{B} \bm{C}^{-1} \bm{B}^\dagger,
\end{equation}
is distributed as $\cw(\bSig/\bm{S},M-r)$, where $\bSig/\bm{S}$ is the Schur complement of the $r\times r$ block in $\bSig$ \cite{ouellette_schur_1981},\cite[p. 539]{rao73}. This result holds only if $M>r$. 

% \vfill
% \newpage
\section{Bayesian inference for spectral quantities}
\label{sec:CPSD_TH_stats_a}

We now use the results from the previous section to perform Bayesian inference of the theoretical distribution underlying a set of observed spectral quantities.\\
Our starting point is the likelihood in Eq.~\eqref{eq:wishart}, which, when multiplied by an appropriate prior distribution $p(\bSig)$, yields the Bayesian posterior for the theoretical CPSD matrix $ \bSig $. Since $\bSig$ is the only free parameter in the sample distribution, this posterior fully captures the statistical information of the stochastic processes under investigation.\\
The key step in this approach is selecting a suitable prior. Before addressing the general case for $p>1$, we begin with the simpler, yet illuminating case of $p=1$, the inference of the PSD of a single stochastic process, which provides valuable insight for the general case.

\subsection{\label{sec:PSD}Inference of the PSD for a single stochastic process}
When $p=1$, Eq.~\eqref{eq:wishart} becomes Eq.~\eqref{eq:directPSD}, and to build a posterior for the PSD $S$ we need a prior $p(S)$. \\
We have considered three options.
\begin{enumerate}
    \item The uniform, noninformative prior $p(S)=\Theta(S)$, with $\Theta(S)$ the Heaviside theta function. With this choice, the posterior distribution of $S$ conditional on the observation of $\Pi$ is
    \begin{equation}
        S\vert \Pi\sim\text{inv}\Gamma(M-1,M\Pi)
    \end{equation}
    with $\text{inv}\Gamma$ the inverse gamma distribution. This posterior is only defined for $M>1$.
    \item Jeffreys noninformative prior \cite{Jeffreys}. Calculating the Fisher information $\mathcal{I}(S)$ from Eq.~\eqref{eq:directPSD}, as prescribed by Jeffreys formula, we get  $p(S)\propto \sqrt{\mathcal{I}(S)} \propto 1/S$, for $S\ge0$. 
    
    As $p(\log(s))=S\, p(S)$, the Jeffreys prior is uniform as a function of $\log(s)$ and corresponds then to a complete lack of prior knowledge even on the order of magnitude of $S$, a rather realistic description of the situation in most cases of noise calibration.  
    
    Note that the main property of the Jeffreys prior is the invariance under reparametrization. Thus, the switch $S\to \log(S)$ does not change the prior probability.

    With the Jeffreys prior:
    \begin{equation}
    \label{eq:invgjef}
        S\vert \Pi\sim\text{inv}\Gamma(M,M\Pi)
    \end{equation}
    \item For comparison we have also considered a prior $p(S)=1/S^2$ that yields  the  posterior 
    \begin{equation}
        S\vert \Pi\sim\text{inv}\Gamma(M+1,M\Pi)
    \end{equation}
\end{enumerate}

The three posteriors above carry some bias. To quantify, it is useful to calculate the  posterior predictive distribution of a further observation $\tilde{\Pi}$, conditional on the past observation $\Pi$, the PDF of which is, by definition:
\begin{equation}
    p(\tilde{\Pi} \vert \Pi)=\int_0^\infty p(\tilde{\Pi}\vert S)p(S\vert \Pi)dS.
\end{equation}

\begin{figure}
    \centering
    \includegraphics[width=1\columnwidth]{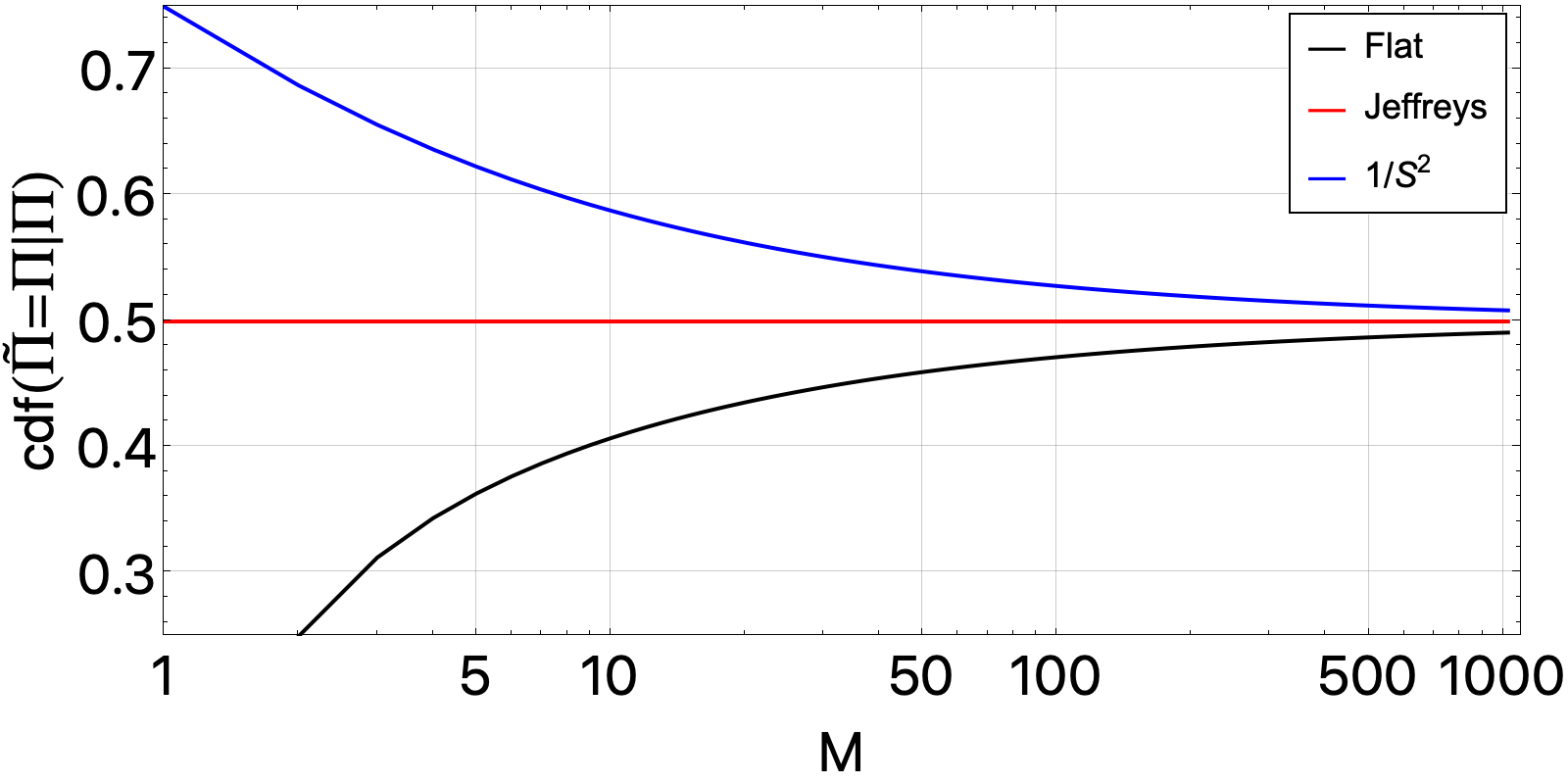}
    \caption{Cumulative distribution function (cdf) of the posterior predictive distribution of a future observation $\tilde{\Pi}$, conditional on the past one $\Pi$, for the three prior options discussed in the text: flat, Jeffreys, and $1/S^2$. The function is  calculated at $\tilde{\Pi}=\Pi$ and plotted as a function of the number of averaged periodograms $M$.}
    \label{fig:CDF}
\end{figure}

The calculation gives   $\tilde{\Pi}\vert \Pi\sim \beta'\left(M,\tilde{M},1,\Pi\right)$ with $\beta'$ the beta prime distribution. The integer $\tilde{M}$ is $\tilde{M}=M-1$, $\tilde{M}=M$, $\tilde{M}=M+1$ for the flat, Jeffreys and  $1/S^2$ priors respectively. 
It seems reasonable that a posterior with minimum bias should assign equal or similar probabilities to future observations larger than the past observation, $\tilde{\Pi}\ge \Pi$, and to those smaller $\tilde{\Pi}\le \Pi$. This means that the cumulative distribution function (cdf) $c(\tilde{\Pi}\vert\Pi)$ should obey $c(\tilde{\Pi}=\Pi\vert\Pi)\simeq 1$. In Fig.~\ref{fig:CDF}, we plot  $c(\tilde{\Pi}=\Pi\vert\Pi)$ as a function of $M$ for the three different priors. The figure clearly shows that, within this definition of bias, the only unbiased choice is the Jeffreys prior. 

\begin{figure}[h]
    \centering
    \includegraphics[width=1\linewidth]{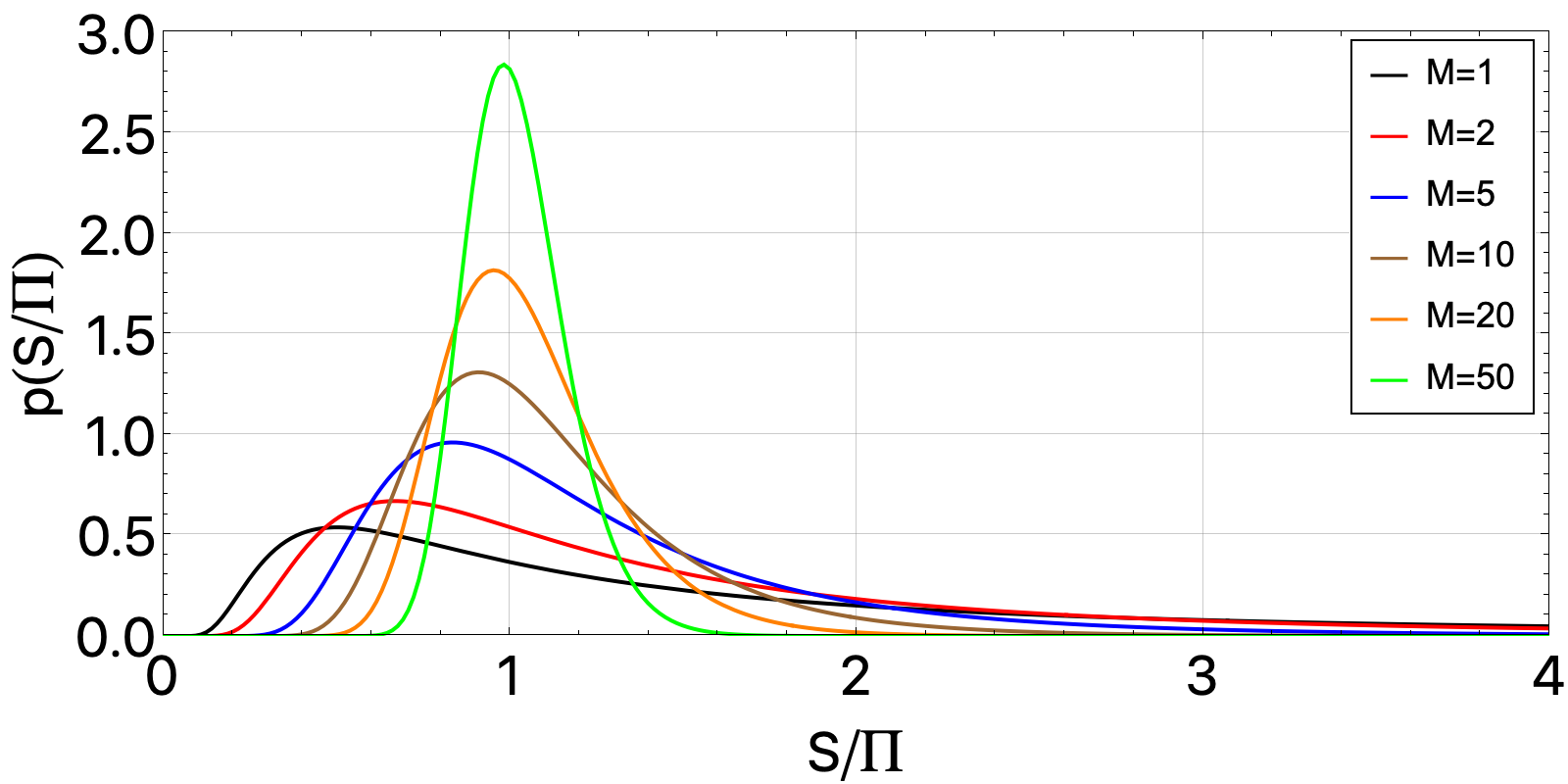}
    \caption{Plot of $p(S\vert\Pi)$ for $S\vert \Pi\sim\text{inv}\Gamma(M,M\Pi)$, for a few different values of $M$. For the sake of clarity, the PDF is for the ratio $S/\Pi$. }
    \label{fig:invgjef}
\end{figure}

In conclusion, given that the Jeffreys prior is unbiased, defined down to $M=1$, invariant under reparametrization, and based on a very realistic assumption about the lack of prior knowledge on the order of magnitude of $S$, we definitely adopt it as the preferred choice.
As a consequence, we adopt the posterior for $S$ in Eq.~\eqref{eq:invgjef}. A plot of the PDF of this posterior for a few choices of $M$ is shown in Fig.~\ref{fig:invgjef}.

Note that at low values of $M$ the PDF is rather skew,  with rather asymmetric equal probability tails around the median. This shows that a naive use of Gaussian statistics may become highly inaccurate. 

To further illustrate this point, we compare the predictions of the posterior in Eq.~\eqref{eq:invgjef} to those of the simplified, Gaussian-based frequentist method, which defines equal-tail credible intervals for $ S $ as $ S \in \left(\Pi \pm k s_\Pi / \sqrt{M} \right) $, where $ s_\Pi $ is the periodogram sample standard deviation. The parameter $ k $ determines the likelihood $ \ell(k) $ of the interval. Since this method is based on Gaussian statistics, it yields $\ell(1) \approx 0.68$, $\ell(2) \approx 0.95$, and $\ell(3) \approx 0.997$. 
We also include in the comparison a variant of this method that accounts for the fact that, in Gaussian statistics, $ t = (S_t - \Pi) / (s_\Pi / \sqrt{M}) $ follows a Student's $t$-distribution with $ M - 1 $ degrees of freedom. Thus, in the calculation of the equal-tail credible intervals, it would be more accurate to replace  $ -k $ by the ${\ell}/{2}$-quantile of the Student-$ t $ distribution with $ M - 1 $ degrees of freedom, and $ +k $ by the $ (1 - {\ell}/{2}) $-quantile.

To perform the comparison, we have done a simulation. Each trial of this simulation consists of the following steps.
\begin{enumerate}[label=(\roman*)]
    \item We extract  $M$ samples $\Pi_i$ from a $\Gamma(1,1)$ distribution, thus  simulating $M$ periodograms of a process with  true PSD  $S_\text{true}=1$.
    \item From the samples above, we calculate the sample mean $\Pi$ and standard deviation $s_\Pi$.  From these we calculate  the credible intervals with likelihoods $\ell(1)$, $\ell(2)$ and $\ell(3)$. We do this for all three methods: the direct frequentist method, the Student-t variant, and the Bayesian posterior in Eq.~\eqref{eq:invgjef}.  
    \item We check which, if any, of these 9 intervals contains the true value $S_\text{true}=1$.
\end{enumerate}

\begin{figure}[h]
    \centering
    \includegraphics[width=1\linewidth]{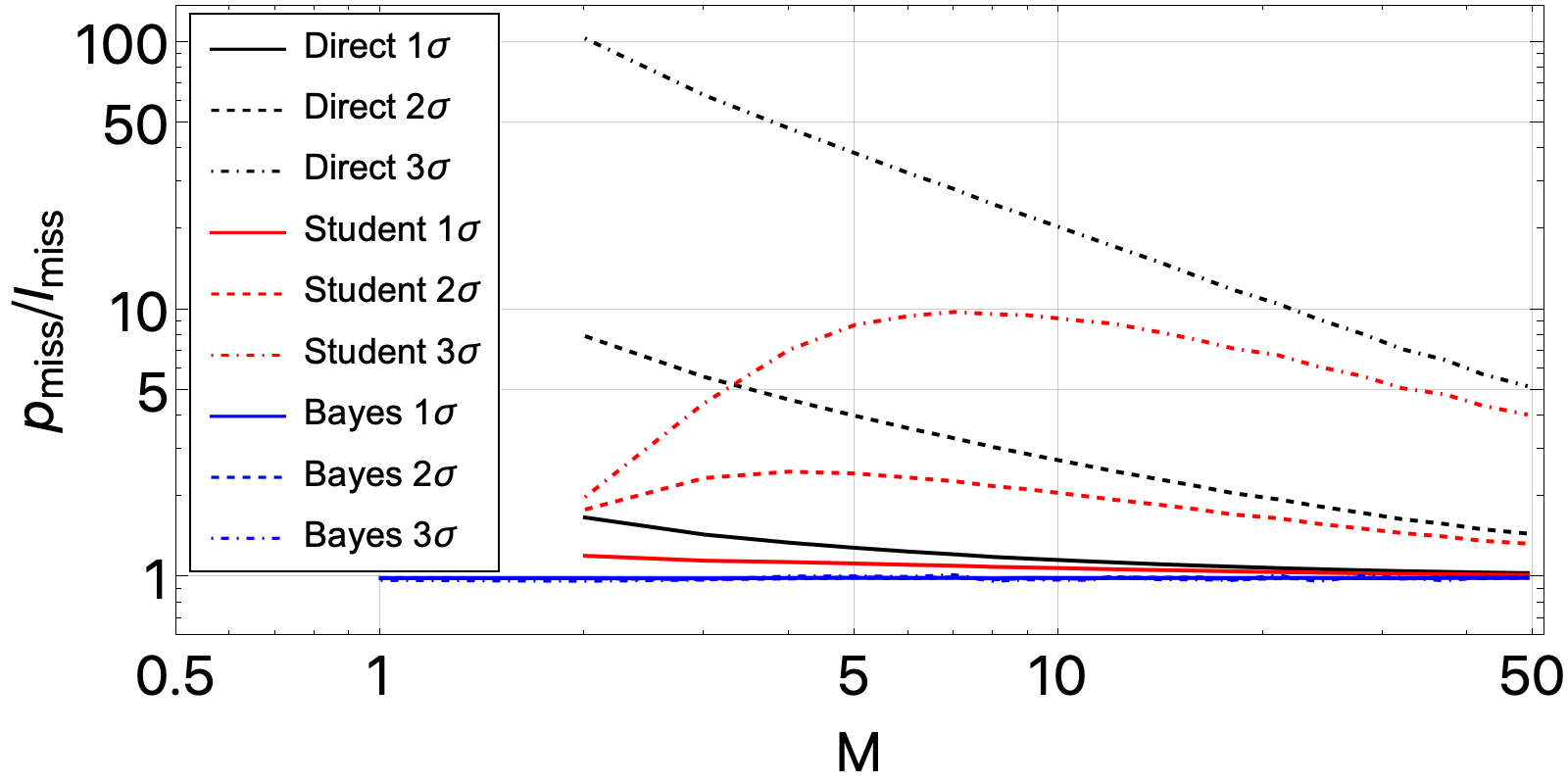}
    \caption{Comparison of the PSD prediction accuracy of the direct frequentist method, its Student-t variant, and the Bayesian posterior in Eq.~\eqref{eq:invgjef}. The plots show the probability $p_\text{miss}$ that the prediction misses the true value, divided by the estimated likelihood of that same event. For each method the simulation has been repeated for equal tail credible intervals with  likelihood $\ell(1) \approx 0.68\;(1\sigma)$ , $\ell(2) \approx 0.95\;(2\sigma)$, and $\ell(3) \approx 0.997\;(3\sigma)$ and as a function of the number $M$ of periodograms in the available sample. The plots for the Bayesian case are barely distinguishable as they are all superimposed on each other at $\simeq 1$, regardless of the value of $M$.}
    \label{fig:prediction}
\end{figure}

By performing a large number of trials, we estimate the probability $p_\text{miss}$ that the true value  $S=1$ is not included within each estimated credible interval, and we compare it with the estimated likelihood of this same event $\ell_\text{miss}=1-\ell$. A consistent estimator should have $p_\text{miss}/\ell_\text{miss}\simeq1$. The results of this simulation are shown in Fig.~\ref{fig:prediction}.

The figure clearly shows that while the Bayesian estimate is consistent and unbiased, the frequentist method may have a probability of missing the true value significantly exceeding the estimated likelihood. The effect increases at low $M$ and may become rather large for  tails beyond the $\ell(1)$ threshold even at $M\simeq 50$. \\
The effect is because the use of Gaussian statistics predicts a credible interval significantly narrower than that predicted by the correct inv$\Gamma$ one. Thus for the same random sample, the  true value may belong to the latter, but  fall outside  the former.

\subsection{\label{sec:CPSDpost} Inference of the entire CPSD matrix}
In the general case $p>1$ some difficulty with the choice of the proper prior for $\bSig$ makes the  spectral  inference more complex. 
Let us start with the basic choice $p(\bSig)=1$ on the positive definite complex matrices domain. With such a choice 
\begin{equation}
    \bSig\vert\bW\sim \cw^{-1}(\bW,M-p)
\end{equation}
with $\cw^{-1}$ the complex inverse Wishart distribution \cite{Shaman_wishart}.

This distribution has a few problems. First, it carries some bias. Though defining what bias is for a matrix distribution may be difficult, it is worth inspecting the posterior predictive distribution of a future observation $\tilde{\bW}$ conditional on the observation of $\bW$.\\ 
We find that $\bW^{-1}\cdot\tilde{\bW}$ is $\bW^{-1}\cdot\tilde{\bW}\sim \mathbb{C}\text{B}_p^{II}(M,M-p)$  with $\mathbb{C}\text{B}_p^{II}(a,b)$  the matrix-variate  type-2 complex Beta  distribution \cite{mathai}. 

Reference~\cite{mathai} shows that $\langle \bW^{-1}\cdot\tilde{\bW}\rangle =\mathbf{I}_p M/(M-2p)$ with $\mathbf{I}_p$ the $p\times p$ identity matrix, while for an unbiased estimation one would expect $\langle \bW^{-1}\cdot\tilde{\bW}\rangle =\mathbf{I}_p $. Note that this mean value bias depends on the number of series considered together and becomes infinite when $M=2p$.

That such dependence of the bias on $p$ is paradoxical is well illustrated by the marginal distribution of the diagonal elements. Indeed, the marginal distribution of $\Sigma_{ii}$, that is the estimate of the PSD of the $i$th time series $S_{i}$, can be calculated  \cite{Shaman_wishart} to be $S_i\vert\Pi_{ii}=\Sigma_{ii}\vert \Pi_{ii}\sim \text{inv}\Gamma(M-2p+1,M\Pi_{ii})$, a distribution only defined for $M\ge 2p$, and different from that one gets by considering the $i$th series alone, $S_i\vert\Pi\sim \text{inv}\Gamma(M-1,M\Pi_{ii})$. \\
Thus, just assuming there are other $p-1$ series that may be correlated with the one under study would change the inferred posterior for $S_{i}$, and would induce a bias increasing with $p$.

The situation is slightly better for the Jeffreys prior. This can be calculated to be \cite{Svensson} $p(\bSig)=\lvert\bSig\rvert^{-p}$ yielding 
\begin{equation}
    \bSig\vert\bW\sim \cw^{-1}(\bW,M).
\end{equation}
With this choice, $\langle \bW^{-1}\cdot\tilde{\bW}\rangle =\mathbf{I}_p M/(M-p)$. Still $S_i\vert\Pi=\Sigma_{ii}\vert \Pi\sim \text{inv}\Gamma(M-p+1,M\Pi_{ii})$, instead of the almost unbiased posterior $S_i\vert\Pi\sim \text{inv}\Gamma(M,M\Pi_{ii})$ that one gets from the Jeffreys prior in the $p=1$ case. Thus, the bias is somewhat reduced but still paradoxically depends on $p$.

The prior that gives the marginal posterior $\Sigma_{ii}\vert\Pi\sim \text{inv}\Gamma(M,M\Pi_{ii})$,  consistent with that estimated from the $i$th series alone and the Jeffreys prior, is $p(\bSig)=\lvert \bSig\rvert^{-2p+1}$.  For this:
\begin{equation}
\label{eq:cpsdpost}
    \bSig\vert\bW\sim \cw^{-1}(\bW,M+p-1)
\end{equation}

Such prior falls within the class on noninformative priors  for $\bQ=\bSig^{-1}$ discussed in \cite{Svensson}, $p(\bQ)\propto \lvert\bQ\rvert^{-K}\etr \left[- \bQ \mathbf{\Lambda}\right]$. Indeed, remembering that the Jacobian of the transformation $\bQ\to\bSig$ is $\lvert \bSig\rvert^{-2p}$ \cite{Shaman_wishart}, this prior corresponds to $K=-1$ and $\mathbf{\Lambda}=0$, that is $p(\bQ)\propto \lvert\bQ\rvert^{-1}$.

Note that for the posterior in Eq.~\eqref{eq:cpsdpost} $\langle \bW^{-1}\cdot\tilde{\bW}\rangle =\mathbf{I}_p M/(M-1)$. As the distribution only holds for $M>1$, and actually the entire multidimensional Bayesian inference only holds for $M\ge p$, the bias remains smaller than $p/(p-1)$ and independent of $p$.\\
Based on the discussion above, we recommend the $p(\bQ)\propto \lvert\bQ\rvert^{-1}$ prior when in need of inferring the whole CPSD at a given frequency.

When only particular functions of the CPSD are needed, as in some of the following sections, the proper priors will be formulated in terms of those functions and not of the whole CPSD.

\subsection{\label{sec:postrho}Inference of the MSC}

We use Eq.~\eqref{eq:directRho} to derive the posterior distribution for the theoretical MSC $|\rho|^2$ (implicitly intended as $|\rho_{ij}|^2$). We adopt a constant prior (over $1\ge|\rho|^2\ge0$) for $|\rho|^2$, as it has a closed form which is not far from the Jeffreys one (see the end of this subsection for details):
\begin{align}
\label{eq:postRhoFlat}
    p(|\rho|^2\big\vert|\hat{\rho}|^2) &=  (M+1)(1-|\rho|^2)^{M}(1-|\hat{\rho}|^2)^{M-2} \nonumber\\
    &\times \frac{{}_2F_1(M,M,1,|\hat{\rho}|^2\,|\rho|^2)}{{}_2F_1(2,2,2+M,|\hat{\rho}|^2)}
\end{align}

\begin{figure}
    \centering
\includegraphics[width=\columnwidth]{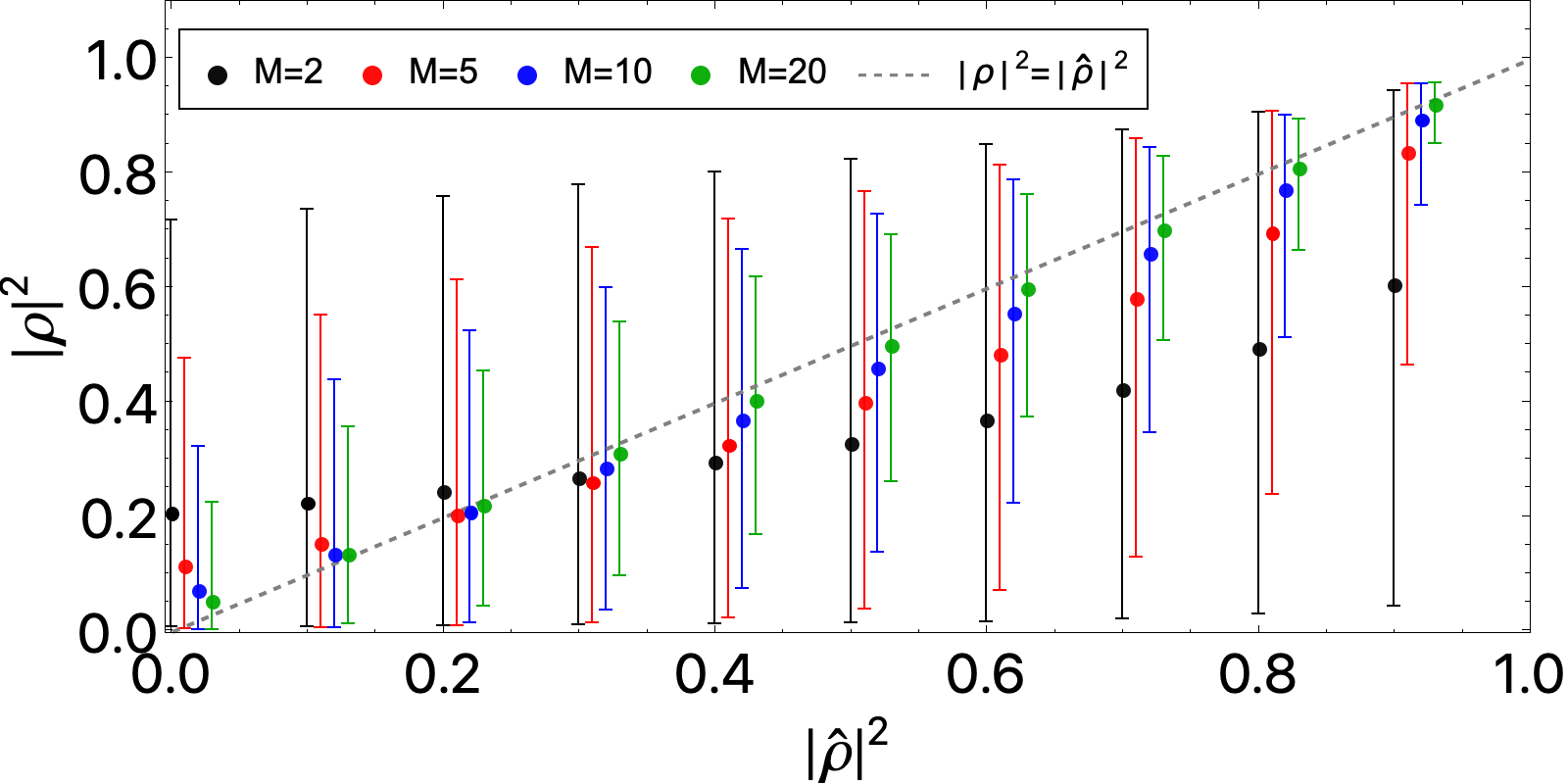}
    \caption{The equal tail, $\ell(2)(\simeq 0.95)$ likelihood credible intervals for MSC (error bars) as a function of the observed value of $\left\lvert\hat{\rho}\right\rvert^2$ and of the number of averaged periodograms $M$. We use a flat prior on $|\rho|^2$. The central dots are the values of the median. The dashed line $\left\lvert \rho\right\rvert^2=\left\lvert\hat{\rho}\right\rvert^2$ is given for reference. For the sake of clarity, for different values of $M$  we plot  $\left\lvert \rho\right\rvert^2$  at slightly shifted values of $\left\lvert\hat{\rho}\right\rvert^2$.}
    \label{fig:rho_theor}
\end{figure}

The MSC is mostly used, within noise characterization, as a diagnostic for the existence of linear correlation between two processes. To illustrate to what extent such a diagnostic parameter is effective, we plot in Fig.~\ref{fig:rho_theor} the $\ell(2)(\simeq 0.95)$ likelihood, equal tail credible interval, and the median predicted by the posterior in Eq.~\eqref{eq:postRhoFlat}. We do that as a function of both $|\hat{\rho}|^2$ and $M$.

\begin{figure}[h]
    \centering
    \includegraphics[width=1\columnwidth]{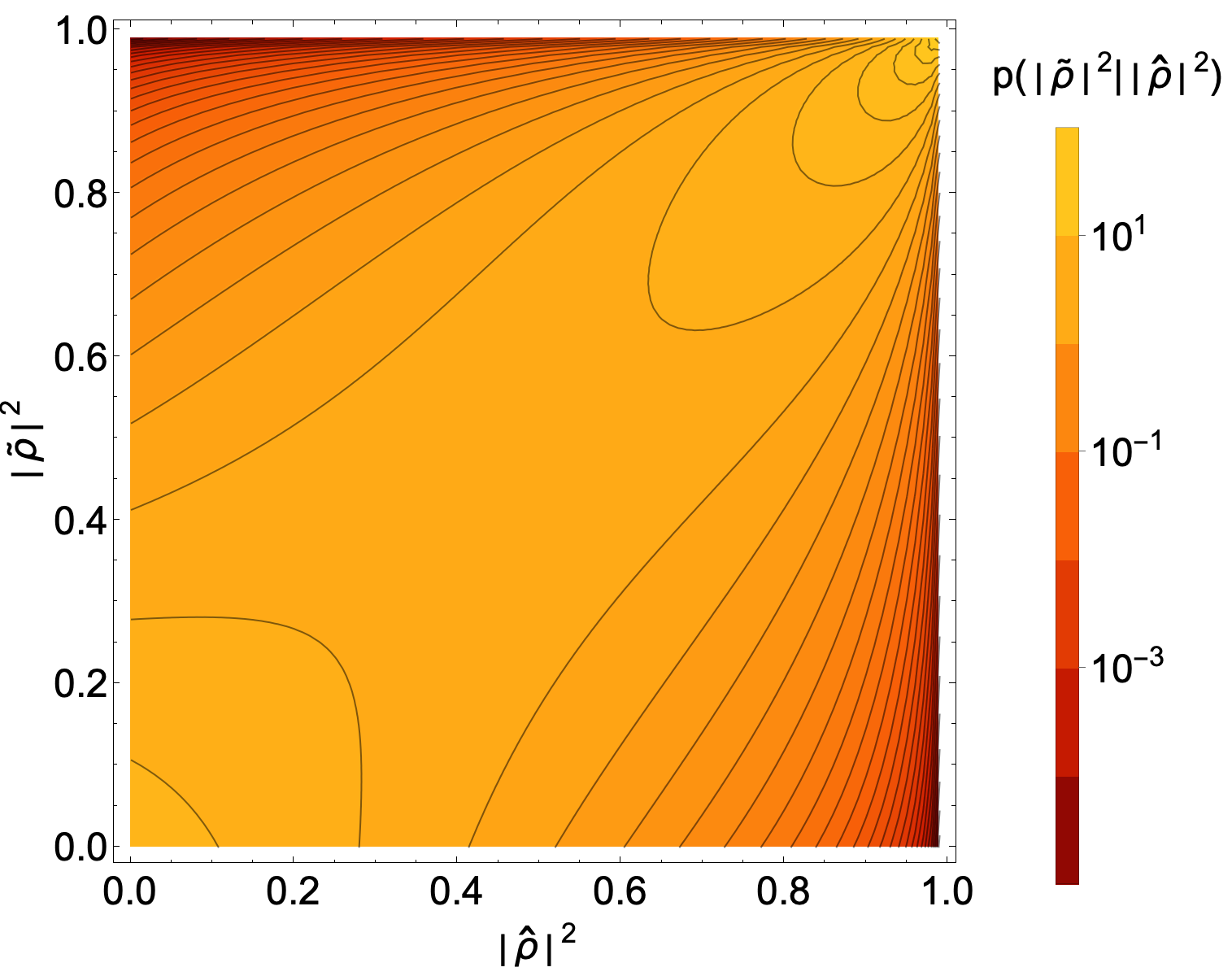}
    \caption{Contour plot of the probability density function $p(\lvert \tilde{\rho}\rvert^2\big\vert \lvert\hat{\rho}\rvert^2)$ of the posterior predictive distribution of a future observation $\left\lvert \tilde{\rho}\right\rvert^2$ conditional on the past observation $\left\lvert\hat{\rho}\right\rvert^2$. The calculation is for $M=5$.}
    \label{fig:contourrho}
\end{figure}

The figure shows that one reaches a reasonable confidence that some correlation exists between the two processes, only when both $M$ and $|\hat{\rho}|^2$ are large enough. For instance, this confidence is never reached  for $M=2$, only if  $|\hat{\rho}|^2\gtrsim 0.6$ for  $M=5$, and even for $M=20$, one would require $|\hat{\rho}|^2\gtrsim 0.2$.\\
Note that the values of the median are always found below the ``unbiased'' line $\left\lvert \rho\right\rvert^2=\left\lvert\hat{\rho}\right\rvert^2$. This bias is only apparent, and in reality, it compensates for the already mentioned significant bias of the sample distribution toward high values. \\
To check this, we have numerically calculated the posterior predictive distribution of a future observation $\left\lvert \tilde{\rho}\right\rvert^2$ conditional on the past observation $\left\lvert\hat{\rho}\right\rvert^2$. We give in Fig.~\ref{fig:contourrho} a contour plot of $p\left(\left\lvert \tilde{\rho}\right\rvert^2\big\vert \left\lvert\hat{\rho}\right\rvert^2\right)$ for $M=5$ that is clearly symmetric around the line $\left\lvert \tilde{\rho}\right\rvert^2=\left\lvert\hat{\rho}\right\rvert^2$, thus showing the lack of real bias of our posterior.

\begin{figure}[h]
    \centering
\includegraphics[width=\columnwidth]{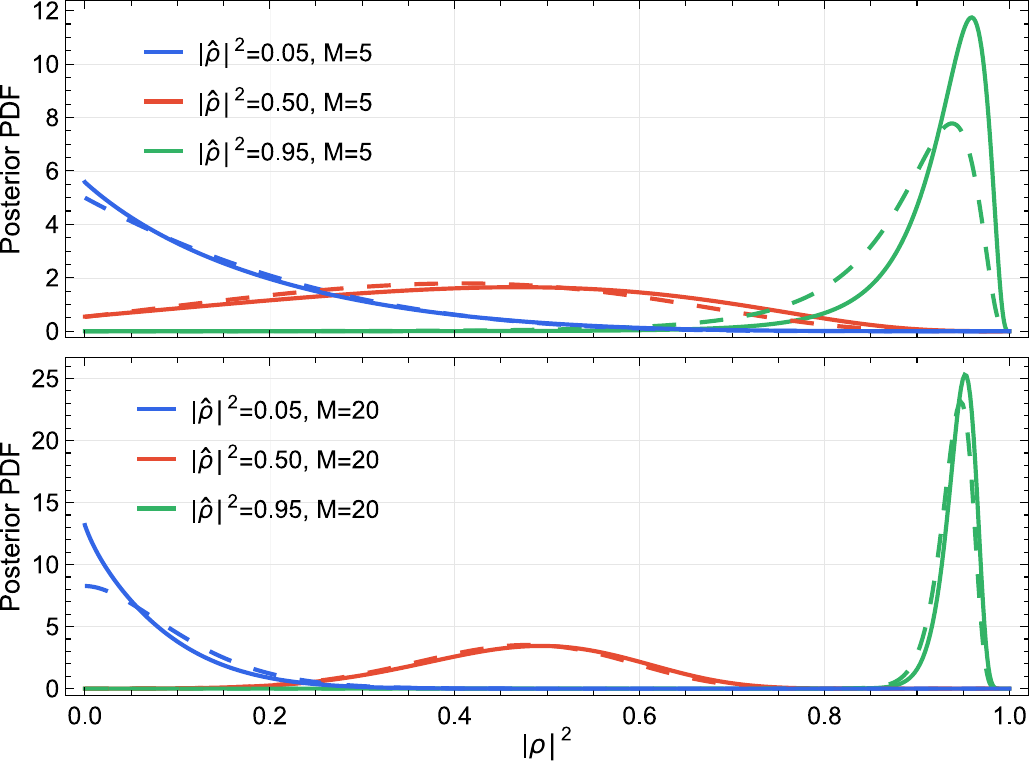}
    \caption{Comparison between posteriors with the Jeffreys prior (solid lines) and the flat prior (dashed lines), with $M=5$ in the upper panel and $M=20$ in the lower panel. Different colors represent different experimental MSCs, $|\hat{\rho}|^2$, as in the legend.}
    \label{fig:postRhoJeffreys}
\end{figure}

To conclude the discussion on the MSC, as anticipated, we show that a flat prior on $|\rho|^2$ does not differ much from the noninformative Jeffreys one. We evaluate numerically the Jeffreys prior $p(|\rho|^2)$, built from the Fisher information as $p(|\rho|^2) \propto |\mathcal{I}(|\rho|^2)|^{1/2}$.
In Fig.~\ref{fig:postRhoJeffreys}, we show the comparison with the flat prior, for a few values of $M$ and $|\hat{\rho}|^2$.

\newpage
\subsection{\label{sec:postR2}Inference of \texorpdfstring{$R^2$}{R2} }

\begin{figure}[h]
    \centering
    \includegraphics[width=1\columnwidth]{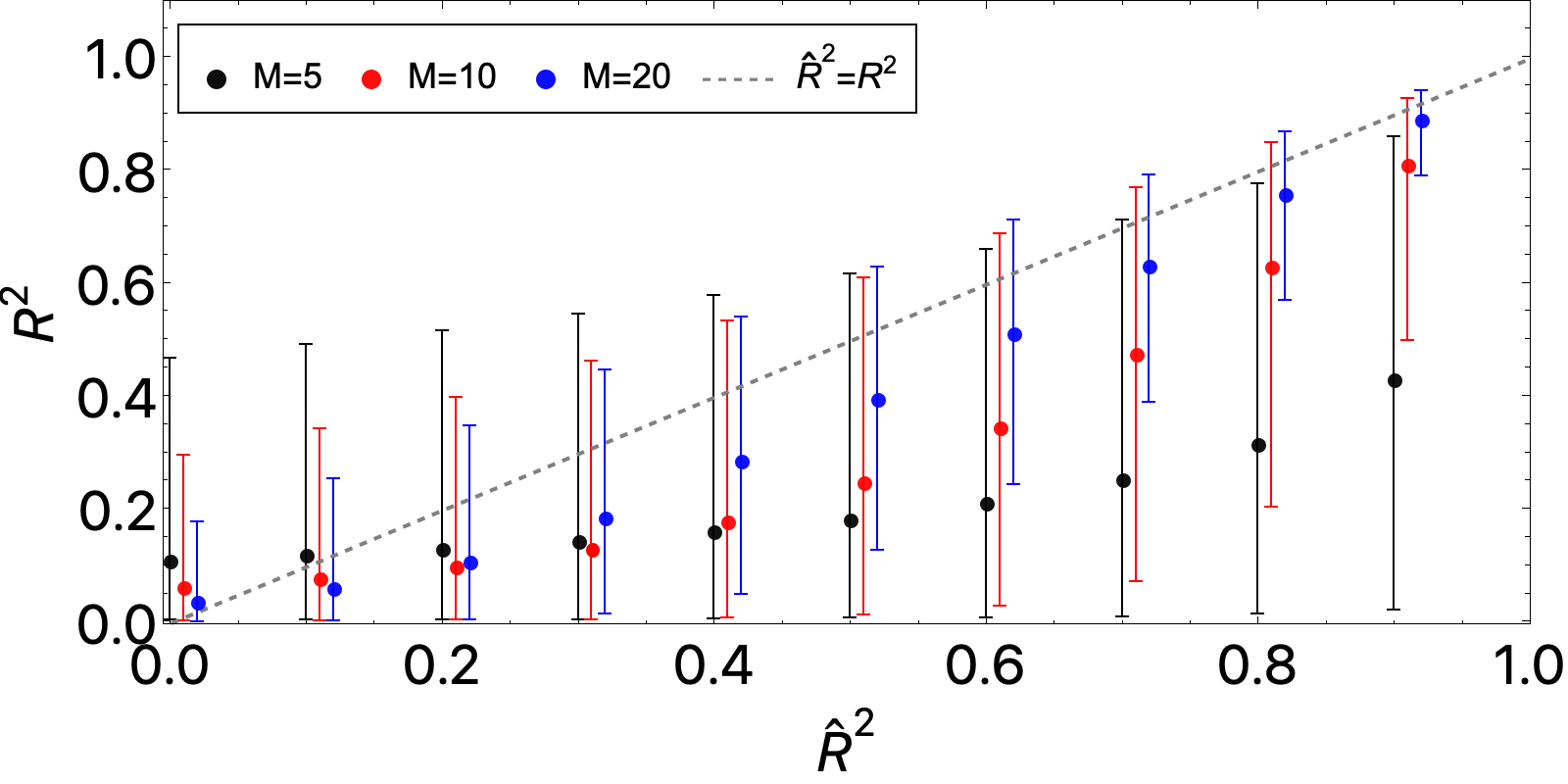}
    \caption{The equal tail, $\ell(2)(\simeq 0.95.5)$ likelihood credible intervals for the multiple coherence  $R^2$ (error bars) as a function of the observed value of $\hat{R}^2$ and of the number of averaged periodograms $M$. We use a flat prior on $R^2$. The calculation is for $p=5$-variate stochastic process. The central dots are the values of the median. The dashed line $R^2=\hat{R}^2$ is given for reference. For the sake of clarity, for different values of $M$  we plot  $R^2$  at slightly shifted values of $\hat{R}^2$.}
    \label{fig:mucorr}
\end{figure}

The case of $R^2$ is very similar to the one in the previous section. The case with Jeffreys prior is not that far from the flat prior, and can be calculated numerically. We give the closed-form posterior with the flat prior:
\begin{align}
\label{eq:postR2}
     p(R^2\big\vert\hat{R}^2) = \,& (M+1)(1-R^2)^{M} \times\nonumber\\& \times\frac{{}_2F_1(M,M,p-1,\hat{R}^2 R^2)}{  {}_pF_q\left(\begin{matrix}
    (1,M,M) \\
    (M+2,p-1)\\
    \end{matrix} ; \hat{R}^2\right)}
\end{align}
where ${}_pF_q$ is the generalized hypergeometric function.\\
As said, $R^2$ is a generalization of MSC for the case $p>2$. We will show in Eqs.~\eqref{eq:Sx0x0},~\eqref{eq:R2_Sx0x0}, that, if $x_1[n]$ is a linear combination of the remaining $p-1$ processes, plus a residual one, then $R^2$ measures the fraction of the total PSD of $x_1[n]$ that is due to the linear combination of the other processes. So ideally, for completely uncorrelated processes, $R^2=0$, while,  in the opposite case of negligible residual, $R^2=1$. \\
Similarly to what we have done for the MSC, to get a sense of the effectiveness of this measure, we plot in Fig.~\ref{fig:mucorr} the $\ell(2)(\simeq 0.95)$ likelihood, equal tail credible interval, and the median predicted by the posterior in Eq.~\eqref{eq:postR2}. We do that as a function of both $|\hat{R}|^2$ and $M$ in the case $p=5$.
The plot shows that also $\hat{R}^2$ carries a very significant bias toward higher values, and the the posterior compensates for such large bias. Again, to conclude that a significant fraction of the noise power in $x_1[n]$ is contributed by the part correlated with the remaining series, one needs comparatively large values both of $M$ and of $\hat{R}^2$.

\newpage
\section{Noise projection and time series decorrelation} 
\label{sec:tsdecorr}

We now consider the case where the $p$-variate stochastic process consists of a ``main'' process $x(t)$ and $r=p - 1$ ``disturbances'' $y_i$, with $1 \le i \le r$, modeled as
\begin{equation}
\label{eq:linmod}
    x(t) = x_0(t) + \sum_{i=1}^r \int_{-\infty}^{+\infty} \alpha_i(t - t')\, y_i(t')\, \mathrm{d}t'
\end{equation}
Here, $x_0(t)$---the ``residual''---is a process independent of the $y_i(t)$'s. We refer to the Fourier transforms of the functions $\alpha_i(t)$, denoted by $\alpha_i(f)$, as the ``susceptibilities''.

Our goal is to estimate the PSD $S_{x_0 x_0}$ of the residual $x_0(t)$, and the susceptibilities $\alpha_i(f)$. This task, often referred to as noise projection or noise decorrelation, is common in what is known as noise hunting, which involves identifying the source of noise in the main data series of a physical apparatus by examining correlations with other independently measured disturbances that may have coupled into the primary measurement. The noise hunting carried out for LPF was no exception, and it required us to develop the approach described below.

We consider two cases:
\begin{enumerate}
    \item When ${\alpha}_i(t)$ is a general function. This allows estimation of both $\alpha_i(f)$ and $S_{x_0 x_0}(f)$ independently at each frequency.
    \item When ${\alpha}_i(t) = \alpha_i \delta(t)$, with $\alpha_i$ constant. In this case, $\alpha_i(f)=\alpha_i$ becomes frequency independent, and thus a global parameter of the estimate, while   $S_{x_0 x_0}(f)$ remains dependent on frequency.
\end{enumerate}

It is convenient for the rest of the discussion to make the following block partition of $\bSig$:
\begin{equation}
\label{eq:blocks}
 \bSig = \left( \begin{array}{c|c}
        S_{xx}& \bS_{xy} \\
        \hline
       \bS_{xy}^\dagger & \bS_{yy}
    \end{array}\right)
\end{equation}
where $S_{xx}$ is just the PSD of $x(t)$, $\bS_{yy}$ is the $r\times r$ CPSD matrix of the disturbances, and $\bS_{xy}$ is the $1\times r$ vector of the CPSD between $x(t)$ and all the disturbances $y_i(t)$. Note that, for clarity,  we have omitted the explicit dependence of all quantities on frequency. We continue to do so in the rest.\\
Within the model in Eq.~\eqref{eq:linmod},
\begin{equation}
\begin{split}
S_{xx}&=S_{x_0x_0}+\sum_{i,j=1}^{r}\alpha_i\alpha_j^*S_{y_i,y_j}= \\&= S_{x_0x_0}+\balp \cdot \bS_{yy} \cdot \balp^\dagger
\end{split}
\end{equation}
where we have  introduced the $r$-long vector $\balp$ with components $\alpha_i$. Furthermore,
\begin{equation}
\label{eq:sxy}
\bS_{xy}=\sum_{i=1}^{r}\alpha_i S_{y_i,y_j}=\balp \cdot \bS_{yy}
\end{equation}
that is $\balp=\bS_{xy}\cdot\bS_{yy}^{-1}$, a relation that will be useful in the following.\footnote{This relation is exactly true only if, at a given frequency $f$,   $\bS_{xy}(f)$ and $\bS_{yy}(f)$ are the exact values at $f$ and not those smoothed over the spectral window (see Eq.~\eqref{eq:wind1}) that we are using here. This spectral smoothing may bias the estimation of $\balp(f)$  should it have a strong dependency on frequency. Such bias can be mitigated by properly reducing the width of the spectral window.}

Note that $S_{x_0x_0}$ is the Schur complement $\bSig/\bS_{yy}$ of  $\bS_{yy}$ in the matrix $\bSig$:
\begin{equation}
\label{eq:Sx0x0}
S_{x_0x_0} = S_{xx}-\bS_{xy}\cdot \bS_{yy}\cdot\bS_{xy}^\dagger= \bSig/\bS_{yy} =1/(\bSig^{-1})_{11}
\end{equation}
and that the  multiple coherence is 
\begin{equation}
\label{eq:R2_Sx0x0}
    R^2 = 1 - \frac{S_{x_0x_0}}{S_{xx}}
\end{equation}
i.e., the fraction of the PSD of $x(t)$ contributed by the disturbances, a useful quantity one wants to estimate.

%The posterior that we found in Eq.~\eqref{eq:postR2} can therefore be used to infer the theoretical contribution of signals to the main measurement.

\subsection{Inference of susceptibilities, residuals, and CPSD of disturbances in the general case}
\label{sec:singlefreq_method}
Our starting point is a key reparametrization of the sample distribution in Eq.~\eqref{eq:wishart}. For the sake of such reparametrization, we need to introduce, in analogy with Eq.~\eqref{eq:blocks}, the block partition of $\bW$ and $\bPi$
\begin{equation}
\label{eq:blockw}
 \bW = \left( \begin{array}{c|c}
        W_{xx}& \bW_{xy} \\
        \hline
       \bW_{xy}^\dagger & \bW_{yy}
    \end{array}\right)=M\left( \begin{array}{c|c}
        \Pi_{xx}& \bPi_{xy} \\
        \hline
       \bPi_{xy}^\dagger & \bPi_{yy}
    \end{array}\right)
\end{equation}

Two functions of $\bW$  that we also need in the following are: 
\begin{enumerate}
    \item The ``observed'' residual noise PSD $\Pi_0$  that we define from
\begin{equation}
\label{eq:so}
   \Pi_0=\frac{1}{M-r}\times\frac{1}{(\bW^{-1})_{xx}}\equiv\frac{1}{M-r}\times W_0
\end{equation}
with $(\bW^{-1})_{xx}$ the upper-left $1\times1$ block of $\bW^{-1}$;
\item The ``observed'' susceptibility vector
\begin{equation}
    \balp_0=\bW_{xy}\cdot\bW_{yy}^{-1}
\end{equation}
\end{enumerate}

In Appendix~\ref{app:DecorrCalculations}, where we show that the sample distribution in Eq.~\eqref{eq:wishart} can be reparametrized as:
\begin{equation}
\label{eq:likelihood2}
\begin{split}
    \hspace{-4.5mm}p(\bW|&S_{x_0x_0},\balp,\bS_{yy}) \propto \frac{1}{S_{x_0x_0}^M} \exp\left[-\frac{W_0}{S_{x_0x_0}}\right]\times  \\
    & \times\exp\left[-\frac{(\balp-\balp_0)\cdot\bW_{yy}\cdot(\balp-\balp_0)^\dagger}{S_{x_0x_0}}\right] \times  \\
    &\times \frac{1}{|\bS_{yy}|^M} \etr \left[-\bS_{yy}^{-1}\bW_{yy}\right]   
    \end{split}
    \end{equation}

Thus, the distribution splits into two independent parts, one depending on $S_{x_0x_0}$ and $\balp$ but not on $\bS_{yy}$, and one that only depends on $\bS_{yy}$. Therefore, if one selects a prior of the kind $p(S_{x_0x_0},\balp)\times p(\bS_{yy})$, then the posterior also splits into the product of the joint posterior for $S_{x_0x_0}$ and $\balp$, with the posterior for $\bS_{yy}$ alone. The estimate of the latter reduce to the estimate of the CPSD that we have already treated. From now on we focus then on the estimate of $S_{x_0x_0}$ and $\balp$ only.

The most realistic, least informative prior for $S_{x_0x_0}$, as for all other PSDs we have met, is again $p(S_{x_0x_0})\propto 1/S_{x_0x_0} $ independently of the value of $\balp$. \\
On the other hand, the components of $\balp$ are, in the language of statics, location parameters. If they can be assumed independent of each other, then, for each of them, the least informative prior is just  $p(\alpha_i)=1$.% over the entire real axis. %commento perchè complesso

From the consideration above, it follows that a sound noninformative   joint prior for $S_{x_0x_0}$, and $\balp$ is $p(S_{x_0x_0},\balp)=1/S_{x_0x_0}$, with which their properly normalized joint posterior becomes:
\begin{equation}
\label{eq:postalfas}
    \begin{split}
    p(&S_{x_0x_0},\balp\vert \bW)=\\
    &= \frac{(W_0)^{M-r}}{\Gamma(M-r)S_{x_0x_0}^{M-r+1}} \exp\left[-\frac{W_0}{S_{x_0x_0}}\right]\times  \\
    &\times\frac{\left\lvert\frac{\bW_{y,y}}{S_{x_0x_0}} \right\rvert}{\pi^r}\exp\left[-(\balp-\balp_0)\cdot\frac{\bW_{y,y}}{S_{x_0x_0}}\cdot(\balp-\balp_0)^\dagger\right].\\
    \end{split}
\end{equation}

This is equivalent to stating that:
\begin{equation}
\label{eq:margso}
    S_{x_0x_0}\left\lvert\right.\bW\sim\text{inv}\Gamma(M-r,W_0)
\end{equation}
and
\begin{equation}
\label{eq:condalpha}
    \balp\left\lvert\right.\bW,S_{x_0x_0}\sim\mathcal{CN}(\balp_0,S_{x_0,x_0}\bW_{yy}^{-1})
\end{equation}
with $\mathcal{CN}(S_{x_0,x_0}\bW_{yy}^{-1},\balp_0)$ the complex, circularly symmetric $r$-variate Gaussian distribution, with covariance matrix $S_{x_0,x_0}\bW_{yy}^{-1}$ and mean value $\balp_0$.\\ 
Note that the marginal distribution of $S_{x_0x_0}$ is already given by Eq.~\eqref{eq:margso} while that of $\balp$ may be obtained by integrating Eq.~\eqref{eq:postalfas} over $S_{x_0x_0}$.
By performing  this integration, we get 
\begin{equation}
    \balp\sim ct_r\left(\balp_0,W_0\bW_{yy}^{-1},M-r\right),
\end{equation}
the latest being the complex multivariate t-distribution for an $r$-long complex vector, with mean value $\balp_0$, scale factor $W_0\bW_{yy}^{-1}$ and $M-r$ degrees of freedom.
This means that  the  real and imaginary parts of $\balp$, recast into the $2r$-long real vector $\balp_R=\left(\begin{matrix}
            \Re\balp \\ \Im\balp
        \end{matrix}\right)$   follow a joint multivariate Student $t_{2r}$ distribution \cite{Multivariate_t} 
    \begin{equation}
     \label{eq:alphacomplex}
        \balp_R\sim
        t_{2r} \left(\balp_{0,R},\,\bm{\Omega},2(M-r)\right),
    \end{equation}
    with $ 2(M-r)$ degrees of freedom, mean value $\balp_{0,R}=\left(\begin{matrix}
            \Re\balp_0 \\ \Im\balp_0
        \end{matrix}\right)$,  and a scale matrix given by
      \begin{equation}
      \label{eq:complexvariance}
        \bm{\Omega}= \frac{1}{2} \, \Pi_0 \,
        \left(\begin{matrix}
            \Re\bW_{yy} & \Im\bW_{yy}\\
           - \Im\bW_{yy} & \Re\bW_{yy}
        \end{matrix}\right)^{-1}
    \end{equation}
    
     From this joint marginal distribution, we also get the marginal distributions of the single components of $\balp_R$ that are univariate $t$ distributions with $2(M-r)$ degrees of freedom \cite{Multivariate_t}, and scale parameter given by the corresponding element in $\bm{\Omega}$.

Note that the covariance of the elements of $\balp_R$, $((M-r)/(M-r-1))\bm{\Omega}$, decreases with decreasing $\Pi_0$, the PSD of residuals. This is expected as, for a given value of the total PSD, a small  $\Pi_0$ implies a large contribution of the disturbances and then a large signal-to-noise ratio for the components of $\balp_R$. 
It is also straightforward to calculate that $\bm{\Omega}\propto M^{-1}$, so that this signal-to-noise ratio, as expected,  also increases with increasing averaging.

To test the validity of our method in a controlled environment, we first studied a simulated case: we generated a set of time series with known correlation, and applied our method to retrieve the known background noise and susceptibilities. This is discussed in Sec.~\ref{sec:simulationsinglefrequency}. We have also used the approach described in this section to decorrelate the effect of the temperature from the acceleration data series of the LPF mission \cite{PhysRevD.110.042004_LPFnoiseperf2024}: details on this are given in Sec.~\ref{sec:LPFtemperature}.

\subsection{Inference in case of additional readout noise}

The model discussed so far assumes the disturbances $y_i(t)$ are measured with negligible readout noise. It is therefore important to consider, before concluding this section, the consequences of applying the method when such  noise is in reality not negligible.

Let us consider first the estimate of $S_{x_0x_0}$. Our method in reality estimates  $1/\Sigma_{11}^{-1}$, whatever the detailed form of $\bSig$ is. Indeed our starting point is $1/W_{11}^{-1}$, whose sampling distribution is $1/W_{11}^{-1}\sim \Gamma(M-r,1/\Sigma_{11}^{-1})$. 
From this, and the distribution in Eq.~\eqref{eq:invgjef}, one can derive  the sampling distribution of our Bayesian estimate for $S_{x_0,x_0}$. We find that this distribution is $\beta'\left(M-r,M-r,1/\Sigma_{11,\text{true}}^{-1}\right)$, a distribution whose median is equal to $1/\Sigma_{11}^{-1}$, and a relative uncertainty that only depends on $M-r$. This confirms that the methods gives an unbiased estimate of $1/\Sigma_{11}^{-1}$.

In the presence of readout noise, $1/\Sigma_{11}^{-1}\neq S_{x_0x_0}$, as  the true form of $\bSig$ is not that in Eq.~\eqref{eq:blocks}. Indeed, in the simplest model of additive noise, the measured disturbance is $y_i(t)+n_i(t)$, with $n_i(t)$ a zero-mean stationary process independent of all the $y_i$'s. In this case, the lower diagonal block of $\bSig$ becomes $\bS_{yy}+\bS_n$ with $\bS_n$ a diagonal matrix whose generic element $S_{n_i,n_i}$ is the PSD of $n_i(t)$. 

Working out the formula for $1/\Sigma_{11}^{-1}$  becomes particularly simple if also $\bS_{yy}$ is diagonal, that is, if the disturbances are mutually uncorrelated. One can readily calculate that in this case
\begin{equation}
    1/\Sigma_{11}^{-1}=S_{x_0x_0}+\sum_{i=1}^r\left\lvert\alpha_i\right\rvert^2 \frac{S_{y_i,y_i}S_{n_in_i}}{S_{y_i,y_i}+S_{n_in_i}}
\end{equation}
with $S_{y_iy_i}$ the PSD of $y_i(t)$. One can recognize that in the limit of dominant readout noise $1/\Sigma_{11}^{-1}\to S_{xx}$. In other words, a dominant readout noise, as expected,  completely obscures any correlation between $x(t)$ and the $y$'s.

Furthermore, within the same simplification of uncorrelated disturbances, the product $\sum_{j=2}^p\Sigma_{1,j} \Sigma_{j,k}^{-1}$, with $k>1$ which, in the noiseless limit, is $(\bS_{xy}\cdot\bS_{yy}^{-1})_{k-1}=\alpha_{k-1}$, becomes instead 
\begin{equation}
 \sum_{j=2}^p\Sigma_{1,j} \Sigma_{j,k}^{-1}=\frac{S_{y_{k-1} y_{k-1}}}{S_{y_{k-1},y_{k-1}}+S_{n_{k-1} n_{k-1}}}\alpha_{k-1}.
\end{equation}

Thus, in the presence of significant readout noise, our method overestimates the PSD of the residuals,  underestimates the absolute value of the susceptibility, and should only be used for an upper limit on $S_{x_0,x_0}$.

\subsection{\label{sec:constantalpha} The case of real frequency-independent susceptibilities}

In many practical circumstances, $\balp$ may be in practice a real frequency-independent vector. If this is the case, $\balp$ becomes a common parameter in the sampling distribution of the $\bW$'s at all frequencies. Thus, to build up a posterior for $\balp$, one needs to consider the joint likelihood of all the $\bW$'s for a given value of $\balp$. We anticipate that, in this case, we do not get a closed form posterior distribution, but a useful form that can be integrated by using the Markov Chain Monte Carlo approach. 

To build this posterior, let us call $\bW_i$ the sample at frequency $f_i$ -- obtained by averaging over $M_i$ periodograms -- and similarly let us indicate the corresponding theoretical quantities with $S_{x_0,x_0,i}$ and $\bSig_{yy,i}$. \\
Let us also assume that $f_i$ and $f_{i+i}$ are sufficiently far apart that $\bW_i$ and $\bW_{i+1}$ may be treated as independent, so that 
the likelihood  of the $\bW_i$ is just the product of their marginal likelihoods $\prod_{i=1}^{N_f} p\left(\bW_i\vert S_{x_0x_0,i},\balp,\bS_{yy,i}\right)$, with $N_f$ the number of considered frequencies.\\
In addition, we assume that $S_{x_0,x_0,i}$ and $\bSig_{yy,i}$ have independent prior distributions,  that the prior for $S_{x_0,x_0,i}$ is, as before, $\propto1/S_{x_0,x_0,i}$, and that the priors for the component of $\balp$ are, again as before, independent and uniform.\\
Using also the fact that, for real $\balp$, Eq.~\eqref{eq:w11a} gives $W'_{x_0 x_0,i}=W_{xx,i}-2\balp\cdot \Re(\bm{W}_{xy,i}) +\balp\cdot\Re(\bm{W}_{yy,i})\cdot \balp$, we get for the logarithm $\Lambda$ of the joint posterior of all parameters
\begin{equation}
\label{eq:multiflik}
\begin{split}
   \Lambda=&-\sum_{i=1}^{N_f}(M_i+1)\log(S_{x_0,x_0,i})-\\
   &-\sum_{i=1}^{N_f}\frac{W_{xx,i}}{S_{x_0x_0,i}}+2 \balp \cdot \sum_{i=1}^{N_f}\frac{\Re(\bW_{xy,i})}{S_{x_0x_0,i}}-\\
   &-\balp \cdot\left( \sum_{i=1}^{N_f}\frac{\Re(\bW_{yy,i})}{S_{x_0x_0,i}}\right)\cdot\balp
   \end{split}
\end{equation}
 plus an independent  term that only contains $\bS_{yy,i}$ and that is not used here.\\
 The posterior in Eq.~\eqref{eq:multiflik} can be used for the MCMC estimate of the parameter posterior distribution.  We have used this method extensively within the data processing of LPF \cite{PhysRevD.110.042004_LPFnoiseperf2024}. To test the validity of our method in a controlled environment, we applied it to the simulated case, as discussed in Sec.~\ref{sec:simulationmultifrequency}.

Before closing this section, it is worth noticing that one advantage of this simultaneous fit to the data at all frequencies, is that one can include also data at very low frequency where the condition $M_i\ge p$ may be violated. This is shown as follows.\\
First, the distribution of a singular  $\bW_i$ obtained from $M_i<p$ periodograms, is the singular complex Wishart distribution \cite{SingularCplxWishart}: 
\begin{equation}
\label{eq:singwishart}
    p\left(\bW_i\big\vert\bSig_i \right) = \frac{\pi^{M_i (M_i-p)}\left|\bm{\Lambda}_i \right|^{M_i-p}}{\widetilde{\Gamma}_p(M_i) \left| \bSig_i \right|^{M_i}} \ \etr \left[- \bSig_i^{-1} \bW_i\right],
\end{equation}
with $|\bm{\Lambda}_i|$,  the product of the nonzero eigenvalues of $\bW_i$.
The dependence of the likelihood in Eq.~\eqref{eq:singwishart} on $\bSig$ and $M$ is the same as that in Eq.~\eqref{eq:wishart}. Thus the difference between the two likelihoods makes no difference for the derivation of the posterior.
Furthermore, the stability of the posterior in Eq.~\eqref{eq:multiflik}, requires that the matrices $\sum_{i=1}^{N_f}\frac{\Re(\bW_{xy,i})}{S_{x_0x_0,i}}$ and $\sum_{i=1}^{N_f}\frac{\Re(\bW_{yy,i})}{S_{x_0x_0,i}}$ are full rank, not the individual $\bW_i$. As the rank of a sum of positive semidefinite matrices is larger than or equal to the maximum rank of the terms in the sum \cite{RankStackOverflow}, it suffices that just one of the $\bW_i$ is of full rank to give full rank to both sums.\\
As, in practice, all the $\bW_i$ above a certain frequency are full rank, the posterior is well defined even if a few terms at the lowest frequency have $M_i<p$.

\section{Test simulations and application to real data}\label{sec:simulationsapplications}

\subsection{\label{sec:simulationsinglefrequency} Test simulation: Single frequency}

We have tested our method in a controlled simulation environment. We have generated a  time series $x(t)=x_0(t)+\sum_{i=1}^3 n_i z_i(t)$, with all series Gaussian and zero-mean, and with  $n_i$ three real coefficients. We have also generated the three ``observed'' disturbances $y_i(t)=h(t)*z_i(t)$, with $h(t)$ the impulse response of a low-pass filter, and with $*$ indicating the time-convolution.

Within this simple model, the susceptibilities become $\alpha_i(f)=n_i/h(f)$ with $h(f)$ the frequency response of the filter, that is the Fourier transform of $h(t)$. The susceptibilities are then complex, frequency-dependent, and noncausal.

\begin{figure}[b]
    \centering
    \includegraphics[width=0.90\linewidth]{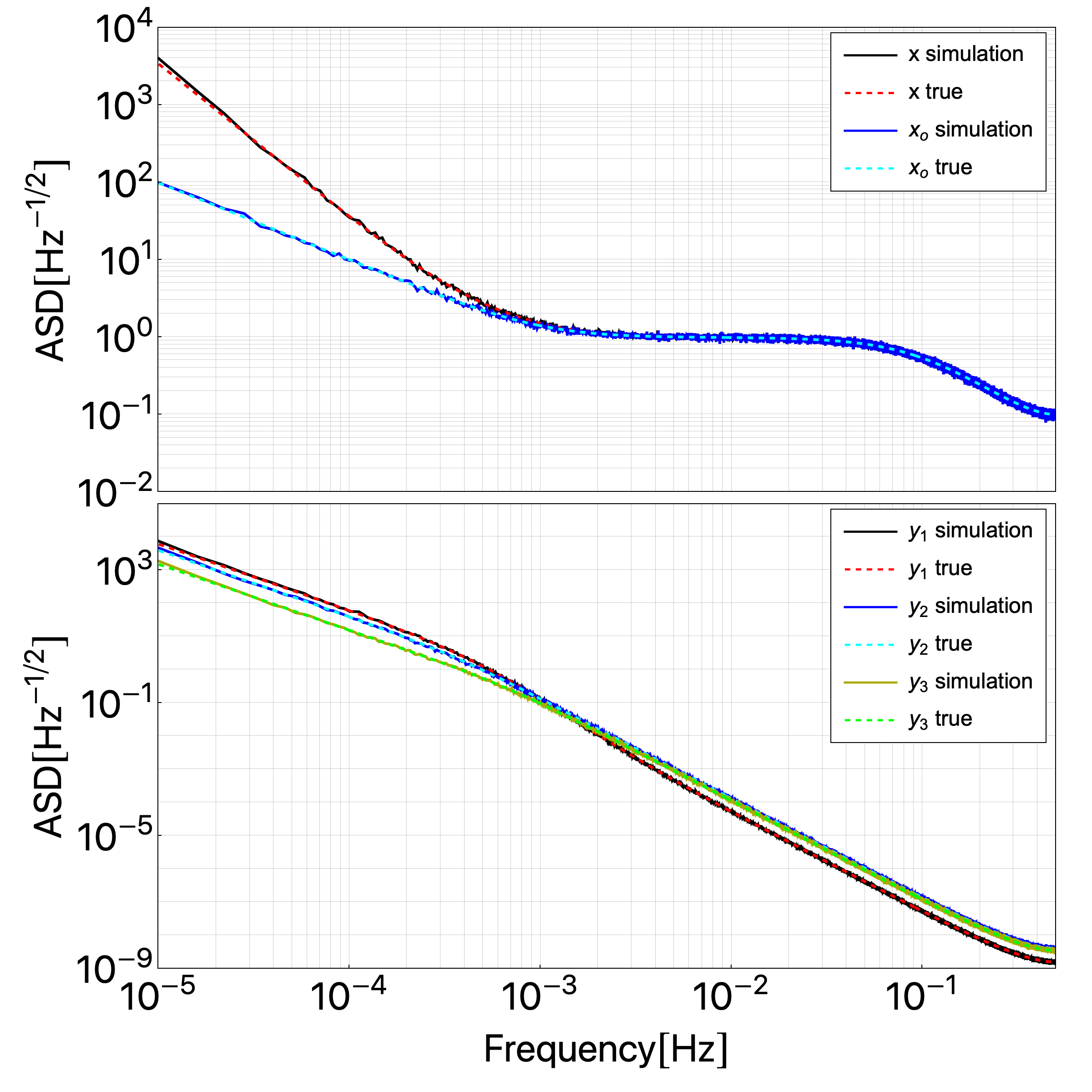}
    \caption{ASD of time series used in the simulation (Welch periodograms). The dashed lines represent the `true values' that is those used to generate the simulated data that have been calculated from Eqs.~\eqref{eq:psdxo} to~\eqref{eq:psdhf}, and from one random extraction of the set of numerical coefficients $n_i$, $c_{lf,i}$, $c_{hf,i}$ and $c_i$. The noisy lines are the averages of the estimated ASD from 100 different simulations generated from the true spectrum above. Time series are sampled with $T=1$ and last $5\times10^5\text{\,s}$. Frequency dependent data partition for periodogram calculation is performed according to the method of Ref.~\cite{PSDnew}. We used the Nuttall four-coefficient minimal-side-lobe spectral window \cite{nuttall1981windows}.}
    \label{fig:simspectra}
\end{figure}

For the simulation, we selected the  PSD of the residuals $x_0$ to consist of a $\propto1/f^2$ low frequency tail merging into a plateau extending up to some double-pole roll-off. More explicitly\footnote{{Note that in this section we show and calculate single-sided PSDs, as this is the standard practice. Discussion, results, and susceptibilities are not affected by this choice.}} (see Fig.~\ref{fig:simspectra}):
\begin{equation}
\label{eq:psdxo}
\begin{aligned}
S_{x_0}(f) &= \left\vert \frac{\left(1-e^{-2\pi f_1 T}\right)}{1-e^{-2\pi f_1 T}e^{-i 2 \pi f T}}  
\frac{\left(1-e^{-2\pi f_2 T}\right)}{1-e^{-2\pi f_2 T}e^{-i 2 \pi f T}} \right\vert^2 \\
&\quad + \left\vert\frac{1-e^{- T/\tau}e^{-i 2 \pi f_0 T}}{1-e^{- T/\tau}e^{-i 2 \pi f T}}\right\vert^2
\end{aligned}
\end{equation}
with  $T=1\text{\,s}$ the sampling time,  $f_1=0.10 \text{\,Hz}$ and $f_2=0.11 \text{\,Hz}$ the two roll-off frequencies, and $f_0=1\text{\,mHz}$  the cross-over frequency between the tail and the plateau.

The disturbances are in the form $z_i(t)=c_{lf,i}z_{lf,i}(t)+c_{hf,i}z_{hf,i}(t)+c_i z_0(t)$, where all the time series on the right hand side are Gaussian, zero-mean and mutually independent, and the coefficient $c_{lf,i}$, $c_{hf,i}$ and $c_i$ are real and randomly selected.

The $z_{lf,i}(t)$ and $z_0(t)$ share the same PSD
\begin{equation}
\label{eq:psdlf}
    S_{lf}(f)=\left\vert\frac{1-e^{- T/\tau_1}e^{-i 2 \pi f_0 T}}{1-e^{- T/\tau_1}e^{-i 2 \pi f T}}\frac{1-e^{- T/\tau_2}e^{-i 2 \pi f_0 T}}{1-e^{- T/\tau_2}e^{-i 2 \pi f T}}\right\vert^2
\end{equation}
with $\tau_1=1.0\times10^5\text{\,s}$ and $\tau_2=1.1\times10^5\text{\,s}$. For $f \gg 1/\tau_1,1/\tau_2$ this PSD amounts to a  $\propto1/f^4$ low frequency tail with unit value at $f=f_0$.

The $z_{hf,i}(t)$ have  PSD
\begin{equation}
\label{eq:psdhf}
    S_{hf}(f)=\left\vert\frac{1-e^{- T/\tau}e^{-i 2 \pi f_0 T}}{1-e^{- T/\tau}e^{-i 2 \pi f T}}\right\vert^2
\end{equation}
again a $\propto1/f^2$ tail with unit value at $f=f_0$. 
The presence of the shared series $z_0(t)$ induces correlation among the $z$'s with CPSD $ S_{z_i,z_j}(f)=c_ic_jS_{lf}(f)$.

All PSDs above must be intended to be zero for $\lvert f\rvert \ge 1/(2 T)$. With this prescription, they can be read as discrete time Fourier transforms of the corresponding discrete time autocorrelation, and their shape allows a straightforward implementation as autoregressive moving average (ARMA) stochastic processes. 
In Fig.~\ref{fig:simspectra}, we illustrate the ASD of the simulated time series. As a first, preliminary assessment, we have checked that all PSDs of the simulated data follow a gamma distribution, as stated in Eq.~\eqref{eq:directPSD}.

As for the filter $h(t)$, its transfer function is 
\begin{equation}
\label{eq:filter}
    h(f)=\frac{1-e^{-2\pi T/\tau_a}}{1-e^{-2\pi T/\tau_a}e^{-i 2 \pi T f}}\frac{1-e^{-2\pi T/\tau_b}}{1-e^{-2\pi T/\tau_b}e^{-i 2 \pi T f}}
\end{equation}
with $\tau_a=2000 \text{\,s}$ and $\tau_a=2001 \text{\,s}$, and zero for $\lvert f\rvert \ge 1/(2 T)$. 

One can recognize the transfer function of a discrete-time two-pole infinite impulse response low-pass filter, which is easily implemented again as an ARMA filter on the discrete time series of the $z$'s.
This model possesses many features of a realistic situation, complex frequency dependence of PSD, cross-correlation among disturbances, high data dynamic range, complex susceptibility etc., to give a meaningful test of the method.

\begin{figure}[t]
    \centering
    \includegraphics[width=\linewidth]{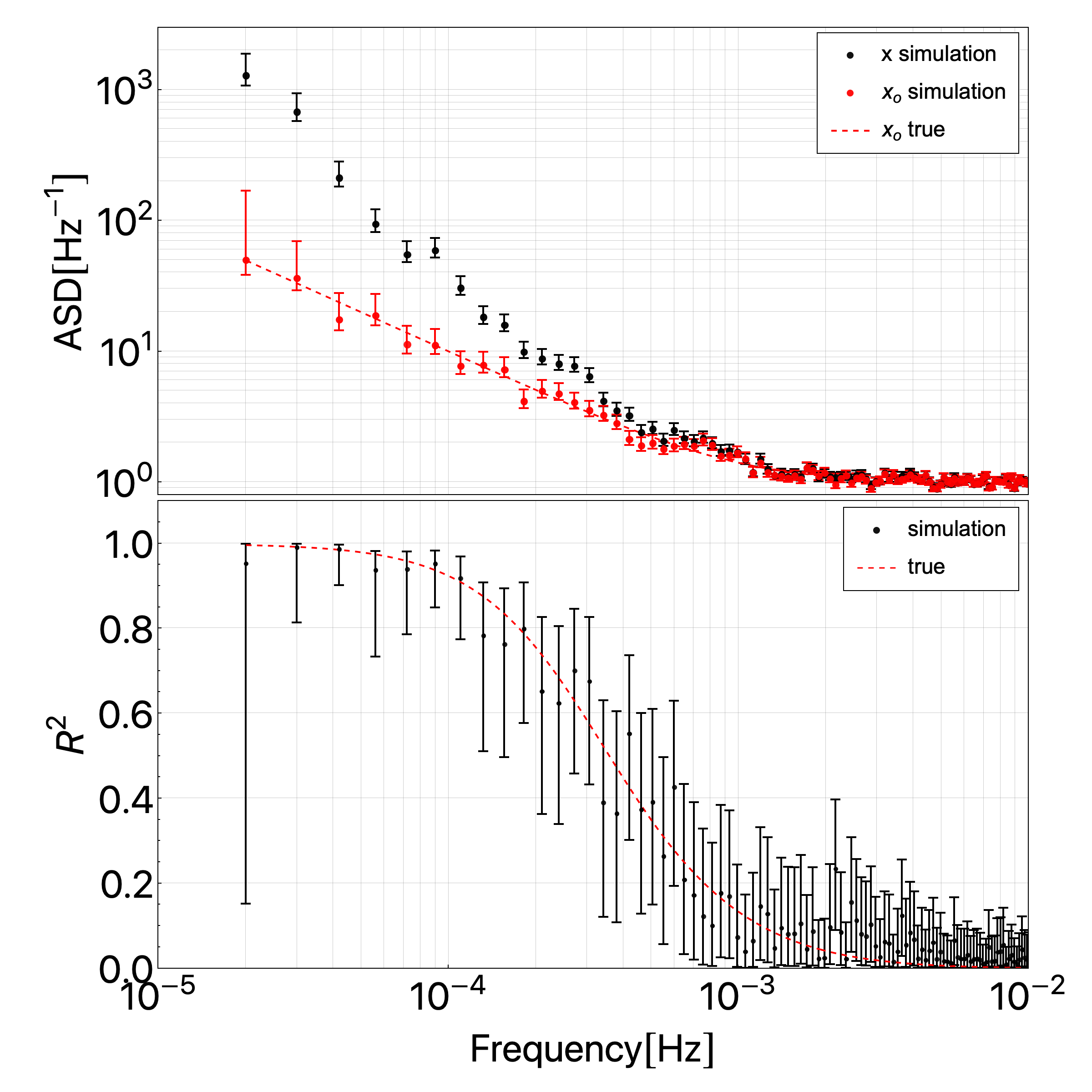}
    \caption{Example of noise decorrelation for a single realization of the multivariate time series $\{x(t),\,y_1(t),\,y_2(t),\,y_3(t)\}$, generated with the same coefficients as in Fig.~\ref{fig:simspectra}. Top panel: black points---posterior ASD of $x(t)$ from Eq.~\eqref{eq:invgjef}; red points---posterior ASD of $x_0(t)$ from Eq.~\eqref{eq:margso}. Points denote medians, with $\ell(1)$ ($\simeq 68.5\%$) equal-tail credible intervals. ASDs are computed with a Nuttall window; frequency spacing yields 10--30\% correlation between adjacent estimates and negligible correlation second-nearest neighbors. The red dashed line shows the true value from Eq.~\eqref{eq:psdxo}. Bottom panel: black points---posterior of $R^2$ from Eq.~\eqref{eq:postRhoFlat}, with medians and $\ell(2)$ ($\simeq 95.5\%$) symmetric credible intervals. The red dashed line shows the true value, $1 - S_{x_0}(f)/S_x(f)$.}
    % \caption{Example of noise decorrelation for one sample of the multivariate time series $\{x(t),\;y_1(t),\;y_2(t),\;y_3(t)\}$ with the same set of numerical coefficient used for the data in Fig.~\ref{fig:simspectra}. Top panel. Black data points: ASD of $x(t)$ estimated using the posterior in Eq.~\eqref{eq:invgjef}. Red data points: ASD of $x_0(t)$ estimated using the marginal posterior in Eq.~\eqref{eq:margso}. Dots represent the medians of the ASD posteriors, while error bars delimit their $\ell(1)$ ($\simeq 68.5\%$ likelihood) equal-tail credible intervals. ASDs have been estimated with the Nuttall window, and the frequency separation is such that nearest neighbors may have a linear correlation in the 10-30\% range. Correlation between the second-nearest neighbors is negligible.  Red dashed line: true value from Eq.~\eqref{eq:psdxo}. Lower panel. Black data points: posterior distribution for the multiple coherence coefficient $R^2$. The posterior is that in Eq.~\eqref{eq:postRhoFlat}, dots are medians, and bars delimit the  $\ell(2)$ ($\simeq 95.5\%$ likelihood) symmetric-tail credible intervals. Red dashed line: corresponding true value for $R^2$ calculated as $1-S_{x_0}(f)/S_x(f)$, with $S_x(f)$ the true spectral density of $x(t)$.}
    \label{fig:decorrspectra}
\end{figure}

In Fig.~\ref{fig:decorrspectra} we apply the decorrelation method described in Sec.~\ref{sec:singlefreq_method} on one example of a $5\times 10^5 \text{\,s}$ multivariate time series generated as described above. 

The figure shows that the method, at least for this example, gives an unbiased estimate of the ASD of the residuals. The ASD indeed fluctuates, within the uncertainties predicted by the posterior in Eq.~\eqref{eq:invgjef}, around the true value in Eq.~\eqref{eq:psdxo}.
The figure also shows, for reference, the estimate of the multiple coherence coefficient $R^2$, from the posterior in Eq.~\eqref{eq:postR2}. The plot indicates that at  $f\simeq 1\text{\,mHz}$, where $M\simeq 30$, the method allows to detect a $\simeq 10\%$ contribution of the disturbances to the total PSD.

\begin{figure}[h]
    \centering
    \includegraphics[width=0.95\linewidth]{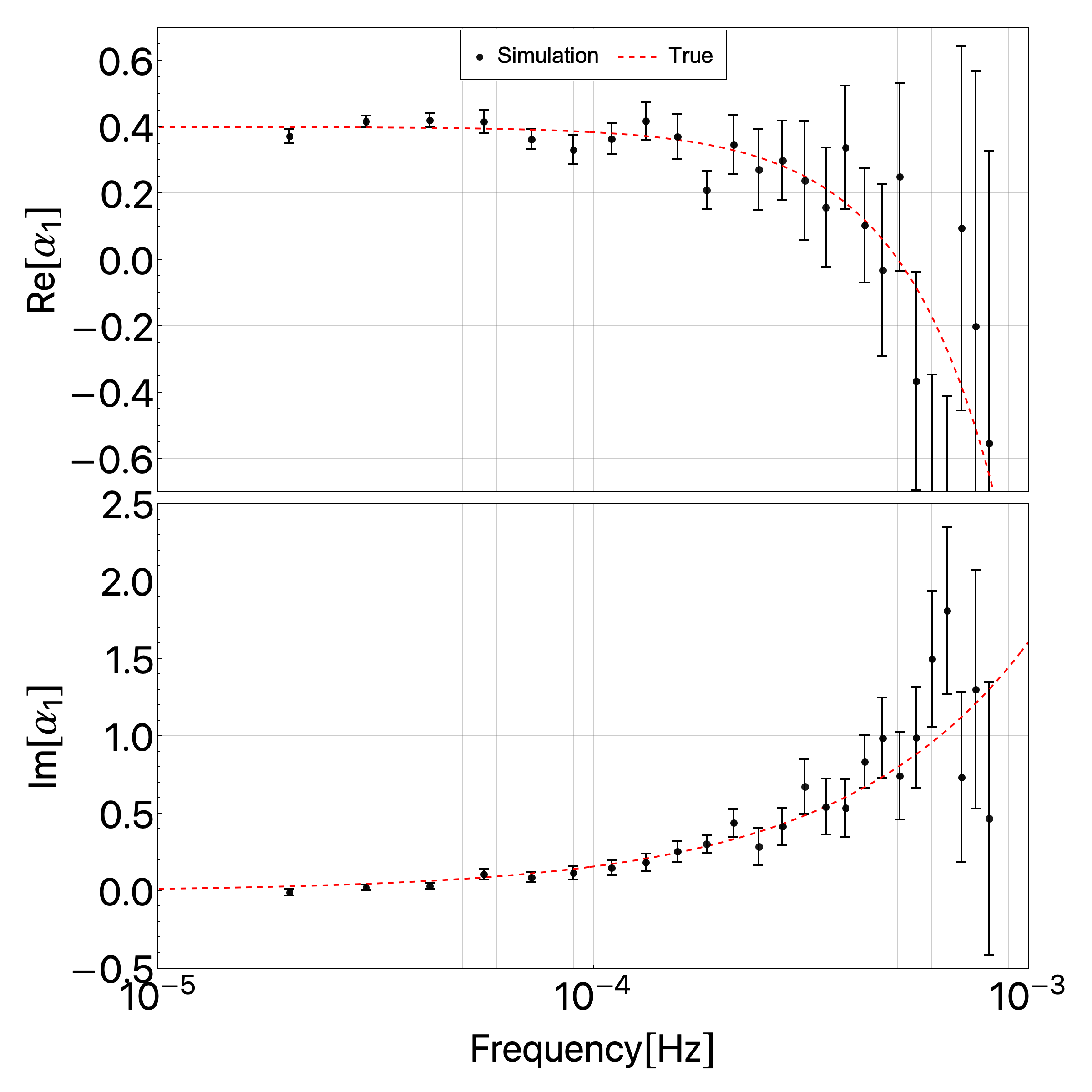}
    \caption{Black data points:  susceptibility $\alpha_1(f)$ of $x(t)$ to $y_1(t)$ for the example data of Fig.~\ref{fig:decorrspectra} from its marginal Student $t$-posterior.  Dots represent the medians of the  posteriors, while error bars delimit their symmetric-tail $\ell(1)$ credible intervals. Red dashed line: true susceptibility value  $n_1/h(f)$, with $h(f)$ from Eq.~\eqref{eq:filter}.  }
    \label{fig:desusc}
\end{figure}

In Fig.~\ref{fig:desusc} we show the estimate of the susceptibility $\alpha_1(f)$ for the same set of data used for Fig.~\ref{fig:decorrspectra}, and compare it with the true value $n_1/h(f)$, with $h(f)$ from Eq.~\eqref{eq:filter}.
The figure is limited to $f\simeq 1\text{\,mHz}$, as above that frequency the susceptibility is in practice compatible with $\alpha_1=0$, as expected from the fact that the contribution of the disturbances to total power becomes undetectable. Within this frequency range the estimate appears unbiased and in agreement with the true value within the uncertainties predicted by the proper marginal $t$-distribution.

\subsection{\label{sec:simulationmultifrequency} Test simulation: Frequency-independent susceptibilities}

\begin{figure}[h]
     \centering
     \includegraphics[width=0.9\linewidth]{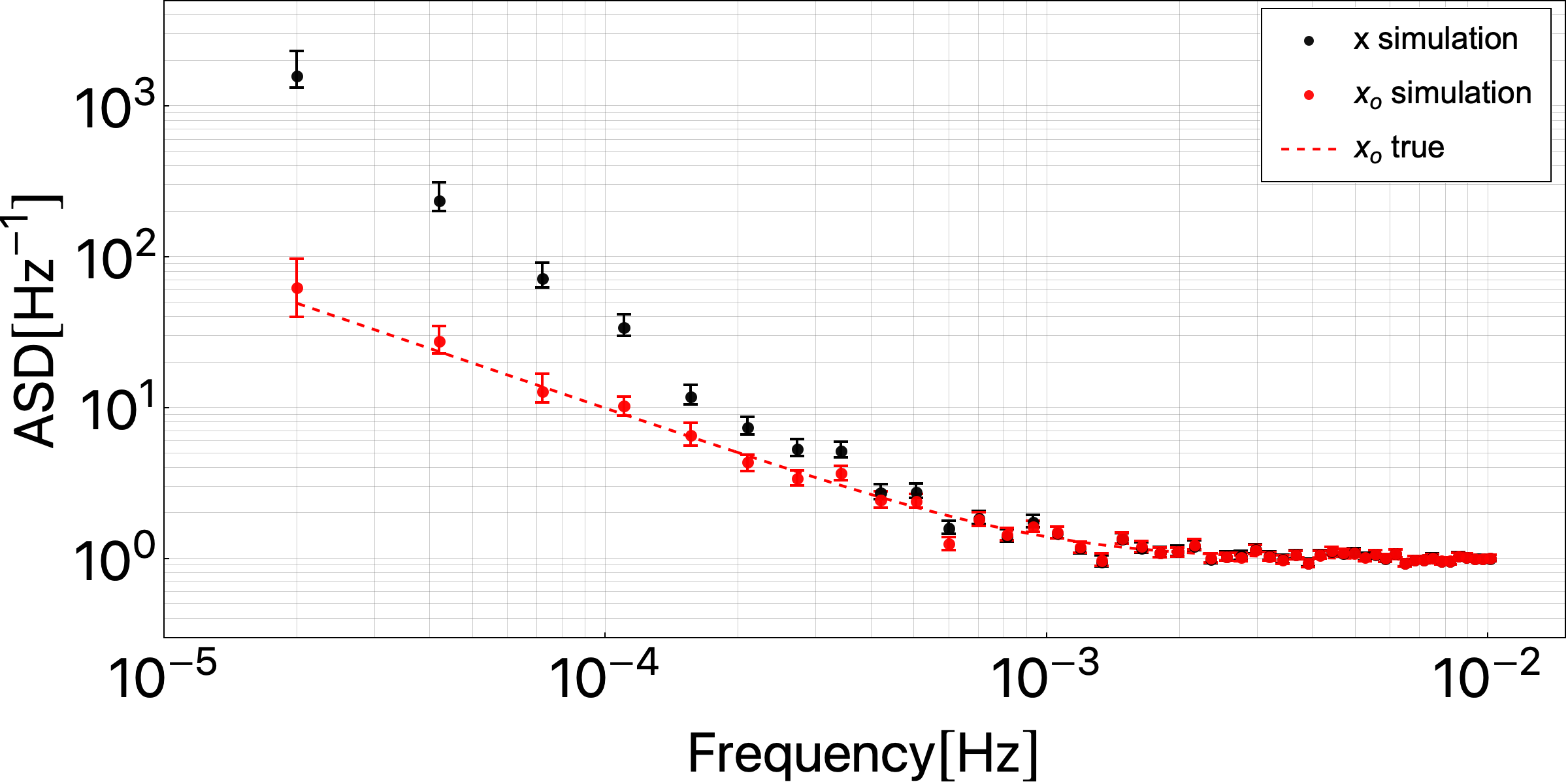}
     \caption{Example of noise decorrelation for one sample of the multivariate time series $\{x(t),\;y_1(t),\;y_2(t),\;y_3(t)\}$ with real frequency independent susceptibilities. Meanings of quantities are the same as those in the upper panel of Fig.~\ref{fig:decorrspectra}. The posterior for the ASD of $x_{0}(t)$ has been obtained with an MCMC integration of the posterior in Eq.~\eqref{eq:multiflik}. Data were taken at every other frequency of the data in Fig.~\ref{fig:decorrspectra}, both to simplify the calculation and to ensure their mutual independence.}
     \label{fig:realsuscdecorr}
 \end{figure}

\begin{figure}[h]
    \centering
    \includegraphics[width=0.9\linewidth]{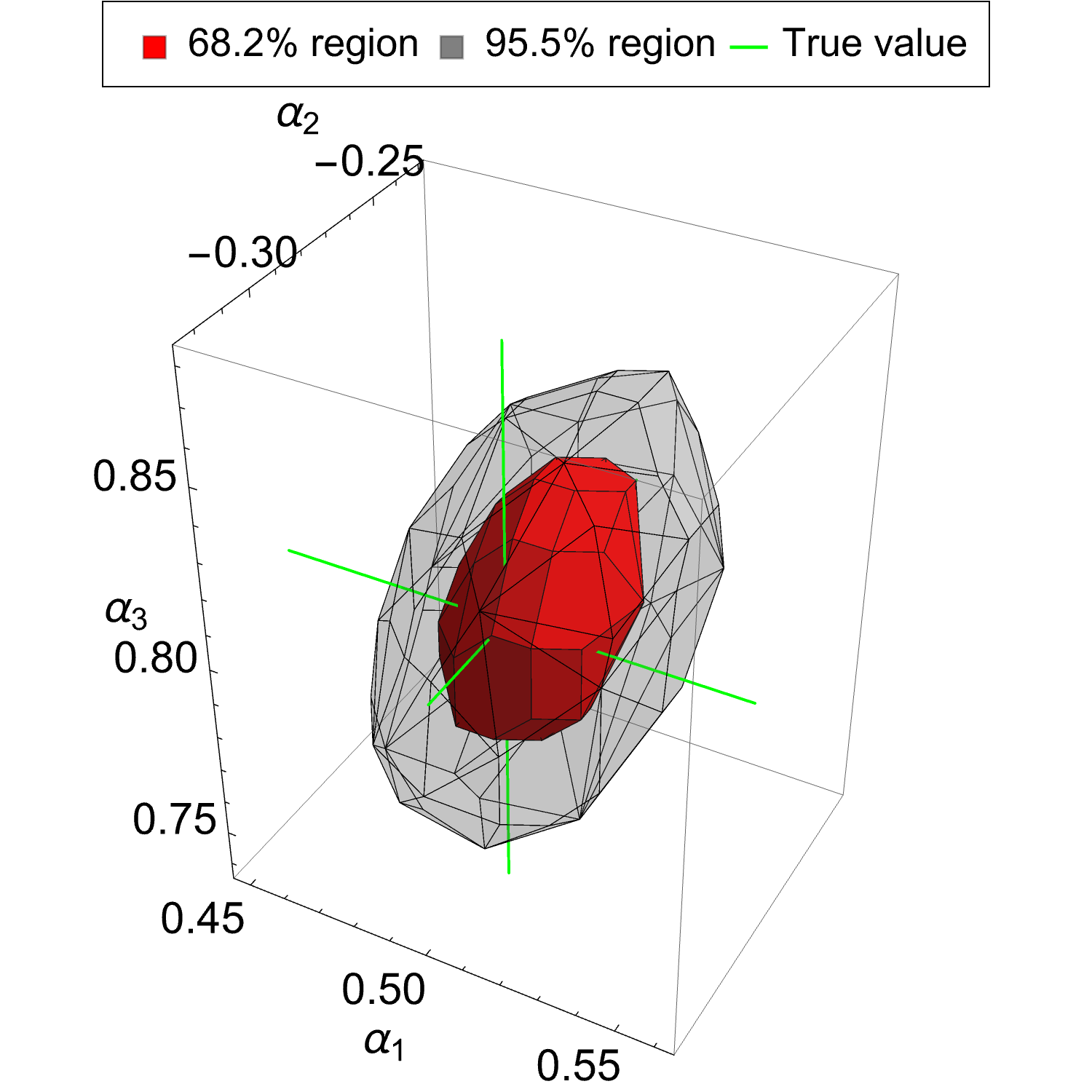}
    \caption{The joint posterior for the three susceptibilities $\alpha_1$, $\alpha_2$, and $\alpha_3$. The red surface delimits a credible region with $\simeq \ell(1)$ likelihood, while the cyan, semitransparent surface delimits a credible region with $\simeq\ell(2)$ likelihood. The green axes cross at the true value $\balp=\mathbf{n}$, with $\mathbf{n}$ the vector with components $n_i$ used in the simulation.}
    \label{fig:realalpha}
\end{figure}

We also tested the method in Sec.~\ref{sec:constantalpha}, applying it to simulations. We choose time series with the same properties as in Sec.~\ref{sec:simulationsinglefrequency}, except that here the disturbances have not been filtered, that is, $y_i(t)=z_i(t)$. With this prescription, the susceptibilities are just the real, frequency-independent numbers $\alpha_i=n_i$. 
The result of such a simulation is presented in Figs.~\ref{fig:realsuscdecorr},~\ref{fig:realalpha}. The figures show again that the method gives a consistent and unbiased estimate of all the  $S_{x_0x_0,i}$ and all the $\alpha_i$.

\subsection{\label{sec:LPFtemperature} Application to real data}
Finally, we apply the method developed in the previous sections to real data from the LPF mission. Over a run lasting 18.5~days, the relative acceleration of two test masses, $\Delta g$, and the average temperature were synchronously measured with a sampling time of \SI{0.1}{s}. The long duration of the measurement allows us to probe frequencies as low as \SI{3.75}{\micro\hertz} with $M=2$. By selecting sufficiently spaced frequencies such that ASD estimates can be treated as statistically independent, we employ the single-frequency approach described in Sec.~\ref{sec:singlefreq_method}. 

\begin{figure}[h!]
    \centering
    \includegraphics[width=1\linewidth]{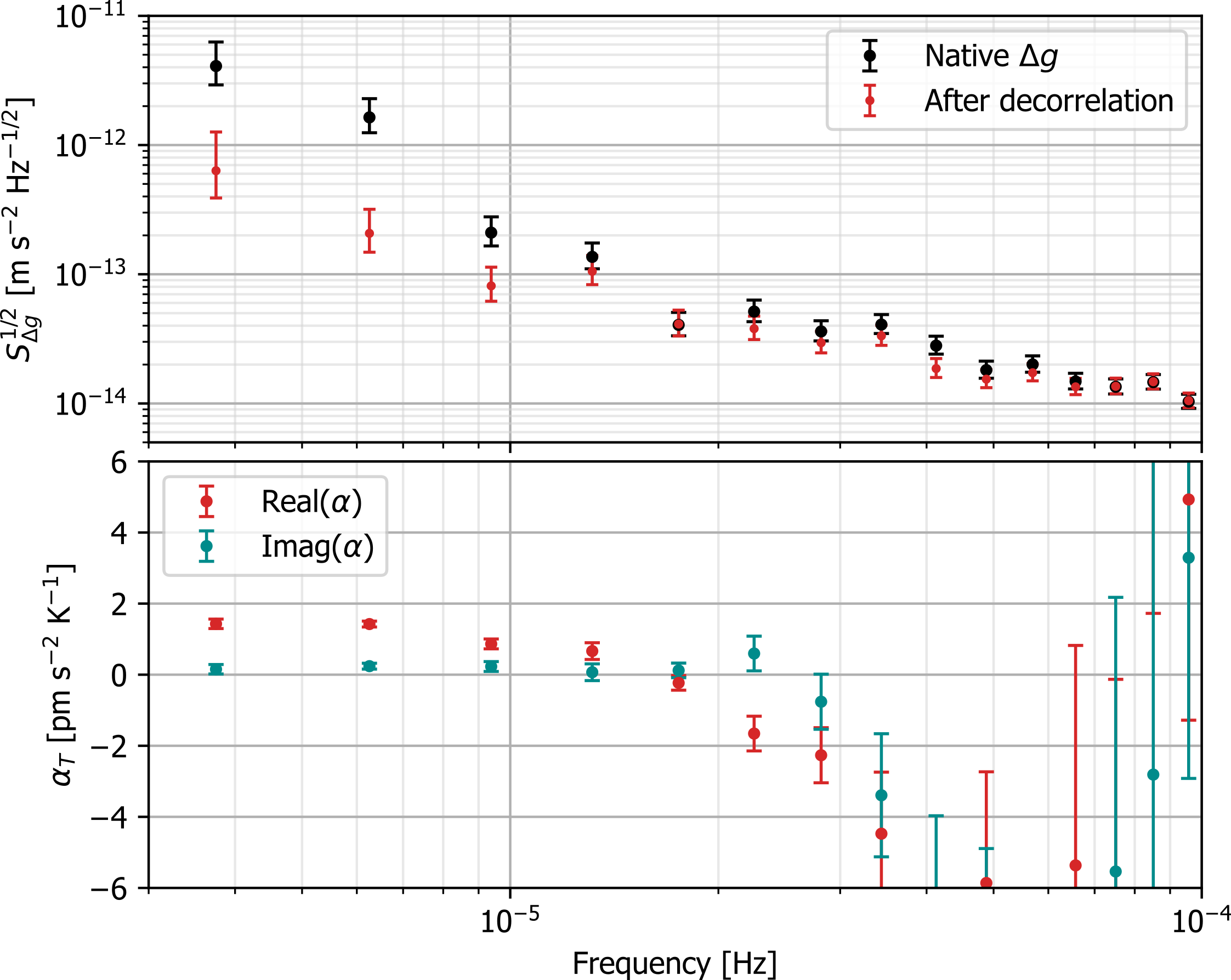}
    \caption{Decorrelation of real data from the LPF mission. Top panel: noise decorrelation, native PSD and projected PSD after the application of our method. Bottom panel: estimate of the susceptibilities, real and imaginary parts.}
    \label{fig:LPF_Dec16_decorrT}
\end{figure}

The results are shown in Fig.~\ref{fig:LPF_Dec16_decorrT}. At \si{\micro\hertz} frequencies, temperature-induced forces dominate the acceleration noise, and the decorrelation procedure proves highly effective. This, in turn, enables a reliable estimation of the susceptibility $\alpha$, indicating that its imaginary part is consistent with zero, and providing a precise measurement of its real part.

% \newpage
\section{Conclusions}
\label{sec:conclusions}
In conclusion, we have presented a set of Bayesian low-bias closed-form posteriors---based on simple and physically meaningful priors---for the most commonly estimated quantities at a given frequency in the spectral analysis of multivariate time series, and in particular in noise projection of physical instruments. 

The distributions of some of these posteriors are available within the main software platforms, which makes the calculation of credible intervals and other statistical quantities particularly simple. For the others, we give the explicit form of the PDF that can be used to numerically calculate the relevant statistical quantities. 
For the reader's convenience, these posteriors are summarized in Table~\ref{tab:all post}.

\renewcommand{\cellalign}{cl}
\renewcommand{\arraystretch}{1.5}
\setlength\extrarowheight{5pt}
\begin{table*}[h]
    \centering
    \begin{tabular}{llccc}
        \multicolumn{5}{c}{\textbf{General Spectral Estimation}} \\
        \midrule
        \textbf{Quantity} & \textbf{Symbol} & \textbf{Posterior} & \textbf{PDF} & \textbf{Section} \\
        \midrule
       Single series PSD & $S$ & \(\displaystyle S\sim\text{inv}\Gamma(M,M \Pi)\) & \(\displaystyle\frac{e^{-\frac{M \Pi }{S}} \left(\frac{M \Pi }{S}\right)^M}{S\; \Gamma (M)}\) & Sec.~\ref{sec:PSD}\\[10pt]
       \hline
      \makecell{Multivariate series\\ CPSD} & $\bSig$ & \(\displaystyle\bSig\sim\cw^{-1}(\bW,M+p-1)\) & \(\displaystyle\frac{\left|\bW \right|^{M+p-1} \ \etr \left[- \bSig^{-1} \bW\right]}{\widetilde{\Gamma}_p(M-p+1) \left| \bSig \right|^{M+2p-1}}\) & Sec.~\ref{sec:CPSDpost}\\[10pt]
      \hline
      Two-series MSC&$\lvert\rho\rvert^2$&--& \makecell[c]{(Flat prior)\\$  (M+1)(1-|\rho|^2)^{M}\times$\\$\times(1-|\hat{\rho}|^2)^{M-2} \frac{{}_2F_1(M,M,1,|\hat{\rho}|^2\,|\rho|^2)}{{}_2F_1(2,2,2+M,|\hat{\rho}|^2)}$}&Sec.~\ref{sec:postrho}\\[10pt]
      \hline
      \makecell{Multiple\\coherence} &$R^2$&--&  \makecell[c]{(Flat prior)\\$(M+1)(1-R^2)^{M}\times$\\$\times\frac{{}_2F_1(M,M,p-1,\hat{R}^2 R^2)}{  {}_pF_q\left(\begin{matrix}
    (1,M,M) \\
    (M+2,p-1)\\
    \end{matrix} ; \hat{R}^2\right)}$}& Sec.~\ref{sec:postrho}\\
        \midrule   
        \multicolumn{5}{c}{\textbf{Noise Projection}} \\
        \midrule
        \textbf{Quantity} & \textbf{Symbol} & \textbf{Posterior} & \textbf{PDF} & \textbf{Section} \\
        \midrule
    \makecell{PSD of residual\\(marginal)} &$S_{x_0x_0}$&\makecell{$S_{x_0x_0}\sim\text{inv}\Gamma(M-r,W_0)$}&\(\displaystyle\frac{e^{-\frac{W_0 }{S}} \left(\frac{W_0 }{S}\right)^{(M-r)}}{S\; \Gamma (M-r)}\) & Sec.~\ref{sec:tsdecorr}\\[10pt]
    \hline
    \makecell{Susceptibilities\\(marginal)}&$\balp_R$& \makecell{$\balp_R \sim $\\ $ t_{2r} \left(\balp_{0,R},\,\bm{\Omega},2(M-r)\right)$}
        & \makecell[c]{$\frac{\Gamma (M) }{\pi ^r (2 (M-r))^r \Gamma (M-r) \|\mathbf{\Omega} |^{1/2}}\times$\\$\times\left(1+\frac{\left(\alpha _R-\alpha _{0,R}\right)\cdot \mathbf{\Omega}^{-1}\cdot \left(\alpha _R-\alpha _{0,R}\right)}{2 (M-r)}\right)^{-M}$}&Sec.~\ref{sec:tsdecorr}\\[10pt]
    \hline
  \makecell{Susceptibilities\\(conditional to $S_{x_0x_0}$)}  &$\balp$&\makecell{$\balp\vert S_{x_0x_0}\sim$\\$\mathcal{CN}\left(\balp_0,S_{x_0x_0}\bW_{yy}^{-1}\right)$}& \makecell[c]{$\frac{\left\lvert\frac{\bW_{y,y}}{S_{x_0x_0}} \right\rvert}{\pi^r}\, e^{-(\balp-\balp_0)\cdot\frac{\bW_{y,y}}{S_{x_0x_0}}\cdot(\balp-\balp_0)^\dagger}$} &Sec.~\ref{sec:tsdecorr}\\
        \bottomrule
    \end{tabular}
    \caption{Summary of closed-form posteriors presented in this paper. For the meaning of the symbols, please refer to the section indicated in the rightmost column.}
    \label{tab:all post}
\end{table*}
\renewcommand{\arraystretch}{1.1}

For the case of noise projection, we have shown with simulations that the method is capable of retrieving, with negligible bias, a residual whose ASD is orders of magnitude smaller than the part due to the measured disturbances, and we have also investigated the robustness of the method in the presence of readout noise in the disturbance measurement.\\
Still on noise projection, in addition to the single frequency closed-form prior, we have also presented a simple likelihood to be used in multifrequency noise projection in case the susceptibilities may be confidently assumed to be real and frequency independent.

We want to stress again that these results originate from the experience of data processing at very low frequency, from \si{\micro\hertz} to Hz \cite{PhysRevD.110.042004_LPFnoiseperf2024}, in which, due to length of the required measurement time, only comparatively few periodograms are available, and should be particularly suitable for any situation in which a similar limitation in number of available periodograms may occur. We have tested the method against controlled simulations, and provided an application to the LPF mission, in the decorrelation of force-inducing thermal effects from the observed acceleration data.\\
Finally, as for almost all commonly used results in data processing, also those presented here are based on Gaussian statistics of the time series under processing. The applicability of the result depends then on how much the data may be safely considered Gaussian.

\section*{Acknowledgments}
This work has been supported in part by Agenzia Spaziale Italiana (ASI), Project No. 2017-29-H.0-2020 ``Attivit\`a per la fase A della missione LISA'', and Project No. 2024-36-HH.0 ``Attivit\`a per la fase B2/C della missione LISA''. The authors would like to acknowledge various useful discussions with all the members of the Trento LISA group.

\section*{Data and Software Availability}
The method presented in this paper is available as a Python package at \cite{Sala_noisefinder_2026}. The simulation data are available from the authors upon reasonable request.

\newpage
\appendix
\section{Derivation of noise-projection posterior.}
\label{app:DecorrCalculations}
Starting from the complex Wishart distribution Eq.~\eqref{eq:wishart}, we derive the joint posterior of the decorrelation parameters in Eq.~\eqref{eq:postalfas}. This was synthetically discussed in   \cite{PhysRevD.110.042004_LPFnoiseperf2024}, and we repeat it here, in more detail, for the reader's convenience.

We start by defining the block matrix $\bU$
    \begin{equation}
        \bU = \left( \begin{array}{c|c}
            1 & \balp \\
            \hline
            \bm{0} & \mathbf{I}_r
        \end{array} \right),
    \end{equation}
where $\mathbf{I}_r$ is the $r\times r$ identity matrix.

$\bU$ performs the  linear transformation $x_0\rightarrow x$, $y_i\rightarrow y_i$. Its inverse  is obtained from $\bU$, by simply replacing $\balp$ with $-\balp$. 

A second important  block matrix is,
\begin{equation}
    \bSig' = \left( \begin{array}{c|c}
        S_{x_0 x_0} & \bm{0}  \\
        \hline
        \bm{0}  & \bS_{yy}
    \end{array} \right) 
\end{equation}
It is straightforward to show that
\begin{equation}
    \bSig = \bU \bSig' \bU^\dagger 
\end{equation}
where  $\bU^\dagger$ is  the conjugate transpose of matrix  $\bU$.

As the determinant of $\bU$ is $\lvert\bU\rvert=1$,  then
\begin{equation}
    |\bSig| = |\bSig'| = S_{x_0 x_0}\,|\bS_{yy}|.
\end{equation}

The exponent in Eq.~\eqref{eq:wishart} can be rewritten in terms of $\bSig'$ as:
\begin{equation}
    \etr\left[-\bSig^{-1} \bW  \right] =  \etr\left[-  \bSig'^{-1} \bW'  \right].
\end{equation}
with $\bW'$ defined as $\bW'=\bU^{-1} \bW (\bU^\dagger)^{-1}$.

As 
\begin{equation}
    {\bSig'}^{-1} = \left(\begin{array}{c|c}
        1/S_{x_0 x_0} & \bm{0} \\
        \hline
        \bm{0}  & \bS_{yy}^{-1}
    \end{array}\right)
    \end{equation}
by using the block decomposition
\begin{equation}
    \bW' = \left(\begin{array}{c|c}
        W'_{x_0 x_0} & \bm{W}'_{x_0y} \\
        \hline
        (\bm{W}'_{x_0 y})^\dagger  & \bm{W}'_{yy}
    \end{array}\right)
\end{equation}
we get
\begin{equation}
\label{eq:expsplit}
\begin{split}
    &\etr\left[-  \bSig'^{-1} \bW'  \right]=e^{-\frac{W'_{x_0 x_0}}{S_{x_0 x_0}}}\etr\left[-\bS_{yy}^{-1}\cdot\bm{W}'_{yy}\right]=\\&=e^{-\frac{W'_{x_0 x_0}}{S_{x_0 x_0}}}\etr\left[-\bS_{yy}^{-1}\cdot\bm{W}_{yy}\right]
    \end{split}
\end{equation}
where, in the last term, we have used the straightforward result $\bm{W}'_{yy}=\bm{W}_{yy}$.\\
$W_{x_0 x_0}'$ in Eq.~\eqref{eq:expsplit} may be written as:
\begin{equation}
\label{eq:w11a}
\begin{split}
     W'_{x_0 x_0}&=W_{xx}-\balp\cdot \bm{W}_{xy}^\dagger -\bm{W}_{xy}\cdot \balp^\dagger+\balp\cdot\bm{W}_{yy}\cdot \balp^\dagger\\
\end{split}
\end{equation}

By introducing 
\begin{equation}
    \balp_0=\bm{W}_{x y}\cdot\bm{W}_{yy}^{-1}.
\end{equation}
we get
\begin{equation}
    \begin{split}
        &\balp\cdot\bm{W}_{xy}^\dagger =\balp\cdot\bW_{yy}\cdot \balp_0^\dagger\\
        &\bm{W}_{xy}\cdot \balp^\dagger=\balp_0\cdot\bW_{yy}\cdot \balp^\dagger
    \end{split}
\end{equation}
so that
\begin{equation}
\begin{split}
     W'_{x_0 x_0}&=W_{xx}+(\balp-\balp_0)\cdot\bm{W}_{yy}\cdot (\balp-\balp_0)^\dagger-\\
     &-\bW_{xy}\cdot\bW_{yy}^{-1}\cdot\bW_{x,y}^\dagger.
\end{split}
\end{equation}
But
\begin{equation}
  =W_{xx}-\bW_{xy}\cdot\bW_{yy}^{-1}\cdot\bW_{x,y}^\dagger=\bW/\bW_{yy}=\frac{1}{(\bW^{-1})_{1,1}},
\end{equation}
thus finally
\begin{equation}
\label{eq:w11}
\begin{split}
     W'_{x_0 x_0}=\frac{1}{(\bW^{-1})_{1,1}}+(\balp-\balp_0)\cdot\bm{W}_{yy}\cdot (\balp-\balp_0)^\dagger
\end{split}
\end{equation}

As $W'_{x_0 x_0}$ is independent of $\bS_{yy}$, the probability density in Eq.~\eqref{eq:wishart} splits in the product of two pieces, one that contains only $\balp$ and $S_{x_0 x_0}$, and one that contains only $\bS_{yy}$. 

Using Eq.~\eqref{eq:w11}, and the definition in Eq.~\eqref{eq:so}, we get:
 \begin{equation}
\begin{split}
    p(\bW|&S_{x_0x_0},\balp,\bS_{yy}) = \frac{\left|\bW \right|^{M-p}}{\widetilde{\Gamma}_p(M) } \times\\
    &  \frac{1}{S_{x_0x_0}^M}e^{\frac{(M-r)\Pi_0}{S_{x_0x_0}}}\times e^{-\frac{(\balp-\balp_0)\cdot\bm{W}_{yy}\cdot (\balp-\balp_0)^\dagger}{S_{x_0x_0}}} \times  \\
    &\times \frac{1}{|\bS_{yy}|^M} \etr \left[-\bS_{yy}^{-1}\bW_{yy}\right]
\end{split}
\end{equation}
This is used for Eq.~\eqref{eq:likelihood2}.

\bibliography{mybibfile}

\end{document}